\documentclass[sn-basic,iicol]{sn-jnl}

\jyear{2022}%

\theoremstyle{thmstyleone}%
\theoremstyle{thmstyletwo}%

\theoremstyle{thmstylethree}%

\usepackage{bm}

\definecolor{codegreen}{rgb}{0,0.6,0}
\definecolor{codegray}{rgb}{0.5,0.5,0.5}
\definecolor{codepurple}{rgb}{0.58,0,0.82}
\definecolor{backcolour}{rgb}{0.97,0.97,0.97}
\lstdefinestyle{mystyle}{
    backgroundcolor=\color{backcolour},   
    commentstyle=\color{codegreen},
    keywordstyle=\color{blue},
    numberstyle=\tiny\color{codegray},
    stringstyle=\color{codepurple},
    basicstyle=\ttfamily\scriptsize,
    breakatwhitespace=false,         
    breaklines=true,                 
    captionpos=b,                    
    keepspaces=true,                 
    numbers=left,                    
    numbersep=1pt,                  
    showspaces=false,                
    showstringspaces=false,
    showtabs=false,                  
    tabsize=1
}
\begin{document}

\title[ ]{A framework for structural shape optimization based on automatic
differentiation, the adjoint method and accelerated linear algebra}

\author*[1]{\fnm{Gaoyuan} \sur{Wu}}\email{gaoyuanw@princeton.edu}


\affil[1]{\orgdiv{Department of Civil and Environmental Engineering}, \orgname{Princeton University}, \orgaddress{\city{Princeton}, \postcode{08540}, \state{NJ}, \country{USA}}}


\abstract{Shape optimization is of great significance in structural engineering, as an efficient geometry leads to better performance of structures. However, the application of gradient-based shape optimization for structural and architectural design is limited, which is partly due to the difficulty and the complexity in gradient evaluation. In this work, an efficient framework based on automatic differentiation (AD), the adjoint method and accelerated linear algebra (XLA) is proposed to promote the implementation of gradient-based shape optimization. The framework is realized by the implementation of the high-performance computing (HPC) library JAX. We leverage AD for gradient evaluation in the sensitivity analysis stage. Compared to numerical differentiation, AD is more accurate; compared to analytical and symbolic differentiation, AD is more efficient and easier to apply. In addition, the adjoint method is used to reduce the complexity of computation of the sensitivity. The XLA feature is exploited by an efficient programming architecture that we proposed, which can boost gradient evaluation. The proposed framework also supports hardware acceleration such as GPUs. The framework is applied to the form finding of arches and different free-form gridshells: gridshell inspired by Mannheim Multihalle, four-point supported gridshell, and canopy-like structures. Two geometric descriptive methods are used: non-parametric and parametric description via B\'ezier surface. Non-constrained and constrained shape optimization problems are considered, where the former is solved by gradient descent and the latter is solved by sequential quadratic programming (SQP). Through these examples, the proposed framework is shown to be able to provide structural engineers with a more efficient tool for shape optimization, enabling better design for the built environment.}




\maketitle
\section{Introduction}\label{sec:intro}
\textit{Form finding} has always been an important topic in architectural and structural design, especially in designing lightweight structures such as concrete shells and gridshells \citep{bletzinger2005computational,adriaenssens2014shell}. There are different interpretations of form finding, and we adapt a broad definition herein: ``form finding is a structural shape optimization process which uses the nodal coordinates as variables to find an optimal structural shape that makes efficient use of material, given external loads and boundary conditions'' \citep{veenendaal2012overview}. The optimal use of material is accomplished when the structural elements are subjected to membrane forces (tension and compression) other than bending \citep{bletzinger2001structural} and thus, the objective is to find a structural geometry that minimizes the bending behavior. Traditionally, form finding was done via physical experiments: Antoni Gaudi and other pioneers applied loads to hanging models with specific boundary conditions to find funicular shapes and then inverted them to obtain the optimized geometry for masonry structures \citep{cuvilliers2020constrained}. In this way, the bending behavior is greatly minimized. However, the physical experiment is time-consuming so form finding is now actively conducted by computational methods. There are generally two computational approaches for form finding, in which one approach is to find an equilibrium-state and the other one is based on optimization theory and finite element analysis (FEA).\par 
The equilibrium-state based approach implements computational methods to simulate the physical experiment of hanging models and it has been extended to support loads other than self-weight. The principle of this method is to simulate a model under loads to find an equilibrium-state that makes the structure fully subjected to membrane forces. Techniques such as Force Density Method (FDM) \citep{schek1974force,cuvilliers2020constrained}, Dynamic Relaxation (DR) \citep{barnes1999form}, Particle-Spring Systems Method \citep{kilian2005particle}, and Combinatorial Equilibrium Modeling (CEM) \citep{ole2016combinatorial,pastrana_constrained_2022} can be used to find such equilibrium-states. A comprehensive review on this approach can be seen in \cite{veenendaal2012overview}. This form finding approach outputs an equilibrium-state and a geometry, but the results are only preliminary and do not inform any actual structural response. For instance, methods like FDM is purely geometric and it is not constructionally practicable \citep{veenendaal2012overview}. For methods like DR, the mass and damping properties used for form finding are fictitious and are not physically meaningful \citep{veenendaal2012overview}. By using this approach, the designers still need to decide the actual structural elements used and conduct further FEA to understand how the structure behaves. \par
The optimization theory and FEA based approach is more generalized form finding methodology. In this approach, a mathematical optimization problem is explicitly formulated and solved by optimization algorithms \citep{bletzinger2001structural}. The objective function to minimize is the total strain energy of the structural system, as minimizing the strain energy is equivalent to minimizing the bending \citep{bletzinger2001structural}. In this approach, one does not simply omit the bending in form finding by using axial-only members. Instead, the designers choose the material, cross-section and the type of elements as they wish and build a finite element model (FEM). The objective function is based on the results of the FEA. As a result, this approach preserves the coherence between structural analysis and optimization, because it considers the properties of the actual structural elements and the FEA results.
While equilibrium-state based form finding is widely used, there exist some drawbacks. First of all, this form finding approach is not informed by the material and cross-sectional properties of the actual structural element, which may result in additional iterations in design. Second, this approach will fail if the structure has cantilever-like boundary conditions because the structural elements used in the form finding process can not resist bending moment \citep{shimoda2014non,ding2017new}. Thirdly, this method is not generalized to the form finding of all kinds of structures and is mainly constrained to the design of lightweight structures. In contrast, optimization theory and FEA based structural shape optimization is capable of resolving the aforementioned issues, being a more generalized and rigorous choice of form finding. Some implementations of this approach can be seen in \citep{ding1986shape,bletzinger2001structural,bletzinger2005computational,uysal2007optimum,pugnale2007morphogenesis,tomas2010shape,ding2017new,rombouts2019novel,xia2019simultaneous,xia2021optimization,san2021shape,meng2022shape}, in which the optimizers range from gradient-based (e.g. gradient descent) to gradient-free (e.g. genetic algorithms) and the optimized structures include bridges, cylindrical shell, domes, hyperbolic paraboloid (hypar), simply-supported free-form shells and free-form canopies. Gradient-based optimization is efficient but calculating the gradients of the objective function and constraints with respect to the design variables is a challenging task. Some work offers insights on how the gradients (i.e. \textit{sensitivities}) are calculated. \cite{ding1986shape} conducted a literature review on shape optimization, in which the analytical sensitivity was derived for shell elements. \cite{uysal2007optimum} implemented finite difference to calculate the sensitivity to obtain the optimum shape for shell structures. \cite{rombouts2019novel} derived the analytical sensitivities of a shape optimization problem for gridshells and compared the analytical results to numerical results. \cite{xia2021optimization} calculated the sensitivity using central difference for a shape optimization subproblem and proposed a strut-and-tie model as an analogy for reinforced concrete. \cite{san2021shape} optimized free-form concrete shells considering material damage based on numerical calculations of sensitivities and they pointed out that the material nonlinearity makes analytical calculation hard to implement. Despite extensive work using analytical or numerical sensitivities, there exist challenges and drawbacks of these methods, which hinders optimization theory based structural shape optimization from being widely accepted in structural design. First, the manual derivation is inefficient and error-prone, which restricts it from being widely used. Moreover, the length and the complexity of the mathematical expression make it hard to be implemented by engineers and architects whose expertise is in structural and architectural design, not mathematics. Second, the derived expression is only applicable to a specific type of structural element so one has to derive different sensitivity expressions for different element types, which will be extremely time-consuming. As for numerical calculation, it suffers from truncation and round-off errors \citep{chandrasekhar2021auto}. Moreover, numerical sensitivity calculation in shape optimization means extra finite element analysis (FEA) needs to be conducted. In contrast to gradient-based methods, gradient-free optimizers like genetic algorithms avoid calculating the gradient but it is usually computationally inefficient and it does not scale well with the complexity of the problem. Therefore, it is of great significance to find an efficient and accurate way to calculate the gradients in structural shape optimization, assisting engineers and designers to find efficient geometries of structures.\par
Automatic differentiation (AD) is an efficient method to calculate the derivatives of functions expressed as computer programs \citep{griewank1989automatic,griewank2008evaluating}. Functions expressed as computer programs, no matter how complicated, can be expressed as a sequence of basic operations and by applying the \textit{chain rule} from calculus, AD can obtain the derivative of the function ``automatically". AD is not numerical, as it gives the exact derivative value to the working precision of the computer; AD is not symbolic, because it does not output an explicit mathematical expression, it only outputs the derivative value when one specifies the variable values at which the derivatives are calculated. AD has been widely used in different fields, such as computational fluid dynamics \citep{bezgin2023jax}, molecular dynamics \citep{schoenholz2020jax}, and machine learning \citep{baydin2018automatic}. In structural optimization, the application of AD can be dated back to the work by \cite{ozaki1995higher}, in which an AD framework based on FORTRAN was implemented to optimize mechanical structures. \cite{espath2011shape} implemented AD to optimize shell structures based on NURBS (non-uniform rational B-spline) description. \cite{norgaard2017applications} introduced the applications of AD in topology optimization. \cite{pastrana_constrained_2022} implemented AD for constrained form finding problems using the CEM method. However, AD is still not widely used in form finding and structural shape optimization, which is partly due to the complexity and the high computational cost of the AD packages previously implemented, as pointed out by \cite{van2005review}. In other words, the power of modern AD software library and hardware acceleration has not been exploited in structural engineering to speed up structural shape optimization. Moreover, only very few work proposed framework for shape optimization using AD, and the existing work does not offer insights on how the structure of computer programs should be configured and optimized. For instance, \cite{espath2011shape} directly applied an AD program called TAPENADE AD \citep{hascoet2013tapenade} for shape optimization but the time scale of AD implementation and the structure of the computer program were not shown.\par
To fill the research gaps identified, here a framework for structural shape optimization based on AD and accelerated linear algebra (XLA) is proposed. The framework is is achieved by the implementation of a high-performance computing (HPC) library in Python called JAX \citep{jax2018github}. The complete codes of the framework are available at \href{https://github.com/GaoyuanWu/JaxSSO}{https://github.com/GaoyuanWu/JaxSSO}. The proposed framework is intended to facilitate the employment of structural shape optimization based on FEA and optimization theory. To obtain the gradient of the objective function (strain energy) with respect to the design variables, the adjoint method \citep{tortorelli1994design} is used to reduce the complexity of computation. By implementing the adjoint method, the gradient computation is decoupled into two sub-tasks: i) calculate the sensitivity of the global stiffness matrix of the structure with respect to the design variables; and ii) conduct FEA. Within the framework, a Python package called JAX-SSO is developed for the first task by using AD. The computation in JAX-SSO is accelerated by a compact programming structure that we proposed and JAX's XLA features. Moreover, JAX-SSO also supports accelerators like GPUs and TPUs to speed up the derivative calculation process, enabling researchers to deal with problems with high complexity. Recently, JAX has been implemented for different optimization problems: \cite{paganini2021fireshape} implemented it for shape optimization problems constrained to partial differential equations (PDEs); \cite{chandrasekhar2021auto} developed a framework for topology optimization using JAX. However, to the best of the authors' knowledge, JAX has not been exploited by structural engineers for gradient-based shape optimization.\par
The goal of this paper is to introduce an efficient framework built on AD, XLA and the adjoint method to facilitate the implementation of the shape optimization method based on optimization theory and FEA. The proposed framework also brings the classic and powerful tool, AD, back to structural engineering community with the assistance of the high-performance computing library, JAX. 
The remainder of this paper is organized as follows. In Section \ref{sec:prob}, the structural shape optimization problem is formulated, followed by a brief introduction of the adjoint method and AD. The features of JAX, the proposed framework and the performance of the JAX-SSO package are introduced in Section \ref{sec:jax-sso}. In Section \ref{sec:test}, the framework is validated and applied to several examples, including the form finding of arches and free-form shells with different boundary conditions. Lastly, the capability and the limitations of the framework are concluded in Section \ref{sec:con}.

\section{Problem formulation}\label{sec:prob}
In structural shape optimization, some widely used objective functions to minimize are: i) the total strain energy of the system, i.e., the \textit{compliance}; ii) the maximum Von Mises stress \citep{ding1986shape}; iii) the \textit{volume displacement} as defined in \cite{robles2001study}; and iv) the \textit{stress levelling} as defined in \cite{bletzinger1993form}. Here we consider the minimization of the strain energy, which is equivalent to maximizing the stiffness and reducing the bending in the structure \citep{tomas2010shape,bletzinger2001structural}. The structural shape optimization problem can then be formulated as follows:
\begin{subequations}\label{eq:nlp}
\begin{alignat}{5}
&\text{minimize} \quad C(\bm{x}) = \frac{1}{2}\int\sigma\epsilon \mathrm{d}V = \frac{1}{2}\bm{f}^\mathrm{T}\bm{u}(\bm{x})  \label{eq:optimization}          \\
&\text{subject to: } \quad \bm{K}(\bm{x})\bm{u}(\bm{x}) =\bm{f}    \quad \label{eq:cons1}\\
&  h_j(\bm{x}) \leq 0    \quad j\in\{1,...,n_{ie}\}  \label{eq:cons2} \\
&  g_j(\bm{x}) = 0    \quad j\in\{1,...,n_{e}\}   \label{eq:cons3}
\end{alignat}
\end{subequations}
where $C$ is the compliance, which is equal to the work done by the external load; $\bm{x} \in \mathbb{R}^{n_d}$ is a vector of $n_d$ design variables; $\sigma$, $\epsilon$ and $V$ are the stress, strain and volume, respectively; $\bm{f} \in \mathbb{R}^n$ and $\bm{u}(\bm{x}) \in \mathbb{R}^n$ are the generalized load vector and nodal displacement of $n$ structural nodes; $\bm{K} \in \mathbb{R}^{6n\times6n}$ is the stiffness matrix; $h_j(\bm{x})$ is the $j$-th inequality constraint out of $n_{ie}$; $g_j(\bm{x})$ is the $j$-th equality constraint out of $n_{e}$. Equation \ref{eq:optimization} is the objective function that one wants to minimize by finding the optimal design variables. Equation \ref{eq:cons1} to \ref{eq:cons3} are the constraints of the problem. The first equality constraint (Equation \ref{eq:cons1}) is necessary for structural shape optimization since it is the governing equation for FEA that any structural system should obey, whereas the other constraints (Equation \ref{eq:cons2}-\ref{eq:cons3}) are optional. In the remainder of this work, we refer the optimization problem without the optional constraints as \textit{unconstrained shape optimization} and the problem with optional constraints as \textit{constrained shape optimization}.\par 
In shape optimization, the design variables $\bm{x}$ determine the geometric form of the structure, which can be i) non-parametric: the design variables are the nodal coordinates of every design node in the system; ii) parametric: the design variables are the parameters or nodal coordinates of control points in a geometry representation function \citep{shimoda2014non} such as B\'ezier surface or Non-uniform Rational B-spline (NURBS) \citep{farin2014curves}. \par 
\subsection{The adjoint method}
For gradient-based optimization, calculating the gradient of the objective function $C(\bm{x})$ with respect to the design variables $\bm{x}$, $\nabla C(\bm{x})$, is of great significance. This process is also known as \textit{sensitivity analysis} and $\nabla C(\bm{x})$ is usually referred as the \textit{sensitivity} of the optimization problem. The basic shape optimization problem without optional constraints is as follows:
\begin{subequations}\label{eq:nconsop}
\begin{alignat}{2}
&\text{minimize}& \quad C(\bm{x})&= \frac{1}{2}\bm{f}^\mathrm{T}\bm{u}(\bm{x}) \label{eq:nconsop_obj}          \\
&\text{subject to: }& \quad \bm{K}(\bm{x})\bm{u}(\bm{x}) &=\bm{f}    \quad \label{eq:nconsop_fea}
\end{alignat}
\end{subequations}
There are two approaches for sensitivity analysis, i.e., to calculate $\nabla C(\bm{x})$: the \textit{direct method} and the \textit{adjoint method} \citep{tortorelli1994design,firl2010optimal}. The direct method is firstly illustrated as follows. We first differentiate the compliance (Equation \ref{eq:nconsop_obj}) with respect to the $i$-th design variable:
\begin{equation}\label{eq:dcdx}
        \frac{\partial C}{\partial x_i} = \frac{1}{2}\frac{\partial \bm{f}^\mathrm{T}}{\partial x_i}\bm{u} + \frac{1}{2}\bm{f}^\mathrm{T}\frac{\partial \bm{u}}{\partial x_i}
\end{equation}
In Equation \ref{eq:dcdx}, the first term is usually easy to obtain as the load $\bm{f}$ is usually independent of the design variables and the displacement $\bm{u}$ can be obtained from finite element analysis (FEA). However, the second term $\frac{1}{2}\bm{f}^\mathrm{T}\frac{\partial \bm{u}}{\partial x_i}$ cannot be obtained in a straightforward way because of $\frac{\partial \bm{u}}{\partial x_i}$. We then differentiate the linear FEA equation (Equation \ref{eq:nconsop_fea}) since $\bm{u}$ is also in this equation:
\begin{subequations}\label{eq:diff_fea}
\allowdisplaybreaks
        \begin{alignat}{3}
        \frac{\partial \bm{K}}{\partial x_i}\bm{u} + \bm{K}\frac{\partial \bm{u}}{\partial x_i} &= \frac{\partial \bm{f}}{\partial x_i}\\
         \bm{K}\frac{\partial \bm{u}}{\partial x_i} &= \frac{\partial \bm{f}}{\partial x_i} - \frac{\partial \bm{K}}{\partial x_i}\bm{u} \label{eq:kdudxi}
        \end{alignat}
\end{subequations}
The Equation \ref{eq:kdudxi} is in a form of a system of linear equations, and if this equation is solved, we can obtain $\frac{\partial \bm{u}}{\partial x_i}$. Equation \ref{eq:kdudxi} can be solved by inverting $\bm{K}$ directly or via methods like LU decomposition, which has a time complexity of approximately $\mathcal{O}(n^3)$. Thus, it is not ideal to solve Equation \ref{eq:kdudxi} for $\frac{\partial \bm{u}}{\partial x_i}$ when there are lot of structural nodes within the system.\par
In contrast to the direct method where an extra system of linear equations needs to be solved, the adjoint method avoids solving Equation \ref{eq:kdudxi}. Moving forward from Equation \ref{eq:kdudxi}, $\frac{\partial \bm{u}}{\partial x_i}$ can be expressed as:
\begin{equation}\label{eq:dudxi}
    \frac{\partial \bm{u}}{\partial x_i} = \bm{K}^{-1} (\frac{\partial \bm{f}}{\partial x_i} - \frac{\partial \bm{K}}{\partial x_i}\bm{u})
\end{equation}
Plug Equation \ref{eq:dudxi} into Equation \ref{eq:dcdx}, we have:
\begin{equation}
    \frac{\partial C}{\partial x_i} = \frac{1}{2}\frac{\partial \bm{f}^\mathrm{T}}{\partial x_i}\bm{u} + \frac{1}{2}\bm{f}^\mathrm{T}\bm{K}^{-1}\left[\frac{\partial \bm{f}}{\partial x_i} - \frac{\partial \bm{K}}{\partial x_i}\bm{u}\right]
\end{equation}
Since $\bm{f}^{\mathrm{T}}=\bm{u}^{\mathrm{T}}\bm{K}^{\mathrm{T}}$ and the stiffness matrix is symmetric, we have:
\begin{subequations}
        \begin{alignat}{4}
        &\frac{\partial C}{\partial x_i} = \frac{1}{2}\frac{\partial \bm{f}^\mathrm{T}}{\partial x_i}\bm{u} + \frac{1}{2}\bm{u}^\mathrm{T}\bm{K}^\mathrm{T}\bm{K}^{-1}\left[\frac{\partial \bm{f}}{\partial x_i} - \frac{\partial \bm{K}}{\partial x_i}\bm{u}\right]\\
        &\frac{\partial C}{\partial x_i} = \frac{1}{2}\frac{\partial \bm{f}^\mathrm{T}}{\partial x_i}\bm{u} + \frac{1}{2}\bm{u}^\mathrm{T}\bm{K}\bm{K}^{-1}\left[\frac{\partial \bm{f}}{\partial x_i} - \frac{\partial \bm{K}}{\partial x_i}\bm{u}\right]\\
        &\frac{\partial C}{\partial x_i} = \frac{1}{2}\frac{\partial \bm{f}^\mathrm{T}}{\partial x_i}\bm{u} + \frac{1}{2}\bm{u}^\mathrm{T}\left[\frac{\partial \bm{f}}{\partial x_i} - \frac{\partial \bm{K}}{\partial x_i}\bm{u}\right]\\
        &\frac{\partial C}{\partial x_i} = \frac{\partial \bm{f}^\mathrm{T}}{\partial x_i}\bm{u} - \frac{1}{2}\bm{u}^\mathrm{T}\frac{\partial \bm{K}}{\partial x_i}\bm{u}\label{eq:dcdxi_adjoint}
        \end{alignat}
\end{subequations}
As can be seen in Equation \ref{eq:dcdxi_adjoint}, implementing the adjoint method avoids solving the extra system of linear equations (Equation \ref{eq:kdudxi}) and the sensitivity can be obtained more efficiently. Let $\mathcal{T}(\cdot)$ denotes the computational complexity of an operation, we have the complexity of the direct method as follows when the displacement vector $\bm{u}$ is already known:
\begin{equation}
    \mathcal{T}(\frac{\partial C}{\partial x_i})_{\text{direct}} \sim \mathcal{T}(\frac{\partial \bm{f}^\mathrm{T}}{\partial x_i}) + \mathcal{O}(n^3) + \mathcal{T}(\frac{\partial \bm{K}}{\partial x_i}\bm{u}) \label{eq:com_direct}
\end{equation}
In contrast, the computational complexity of the adjoint method is:
\begin{equation}
    \mathcal{T}(\frac{\partial C}{\partial x_i})_{\text{adjoint}} \sim \mathcal{T}(\frac{\partial \bm{f}^\mathrm{T}}{\partial x_i}) + \mathcal{O}(n) + \mathcal{T}(\frac{\partial \bm{K}}{\partial x_i}\bm{u}) \label{eq:com_adjoint}
\end{equation}
Comparing Equation \ref{eq:com_direct} and \ref{eq:com_adjoint}, it can be seen by applying the adjoint method, the complexity of computation is greatly reduced by approximately $\mathcal{O}(n^3)$, which is great significance for the efficiency of gradient evaluation in shape optimization.
Usually, the generalized load vector $\bm{f}$ is independent of the design variables and the sensitivity in Equation \ref{eq:dcdxi_adjoint} can be further simplified to:
\begin{equation}\label{eq:dcdxi_simplified}
    \frac{\partial C}{\partial x_i}=-\frac{1}{2}\bm{u}^\mathrm{T}\frac{\partial \bm{K}}{\partial x_i}\bm{u}
\end{equation}  
In the remainder of this work, we assume the load vector $\bm{f}$ remains unchanged during the shape optimization process. By using the adjoint method, we can decouple the sensitivity calulcation into two parts: i) obtain the derivatives of the global stiffness matrix with respect to the design variables, $\frac{\partial \bm{K}}{\partial \bm{x}}$; and ii) conduct FEA to obtain the displacement vector $\bm{u}$ and calculate the sensitivity by Equation \ref{eq:dcdxi_simplified}.
\subsection{Unconstrained shape optimization} \label{USO} 
With the sensitivity term derived using the adjoint method, the unconstrained shape optimization problem can be solved by gradient descent by iteratively updating the design variable as follows:
\begin{equation}\label{eq:gd}
    x_i^{k+1} = x_i^k - t_i^k\cdot\frac{\partial C}{\partial x_i^k}
\end{equation}
where $k$ is the iteration step and $t_i^k$ is the step size (a.k.a learning rate). The step size can be a fixed value. However, it cannot be too big otherwise the optimization will diverge, and it can not be too small since an extremely small step size is slow. Step size can also be updated during each iteration by methods like backtracking line search.
\subsection{Constrained shape optimization}\label{CSO} 
In order to achieve other properties of structures besides low strain energy, constrained shape optimization can be conducted. The constraints can be applied to the nodal coordinates, the mass of the structure, the maximum stress in structural elements, the height of the structure, etc. Mathematically, constrained shape optimization is equivalent to constrained nonlinear programming (NLP) problems and can be solved by methods such as various sequential quadratic programming (SQP) \citep{gill2012sequential,boggs1995sequential} methods, including sequential least-square programming (SLSQP) \citep{kraft1988software}. Note that SQP is not a single algorithm, but it is rather a concept that inspires various algorithms. SQP methods are gradient-based and the general steps of SQP methods are illustrated are follows. In SQP, the design variable $\bm{x}^k$ is updated iteratively at each iteration:
\begin{equation}\label{eq:sqp_update}
    \bm{x}^{k+1} = \bm{x}^k + \bm{d}^k
\end{equation}
where $\bm{d}^k$ is the search direction at the $k$-th iteration. In unconstrained optimization by gradient descent, the search direction is essentially the gradient of the objective function as there is not constraint on the design variables. However, in constrained optimization, the search direction $\bm{d}^k$ is found by solving a quadratic programming subproblem to maintain the balance between finding the minimum and keeping the solution feasible. An approximation to the objective function $C(\bm{x})$ can be expressed as follows at an iterate $\bm{x}^k$ via a second order Taylor expansion:
\begin{multline}\label{eq:talor_obj}
    C(\bm{x}) \approx C(\bm{x}^k) + \nabla C(\bm{x}^k)^{\mathrm{T}}(\bm{x}-\bm{x}^k) +\\
    \frac{1}{2}(\bm{x}-\bm{x}^k)^{\mathrm{T}}\mathbf{H}_C(\bm{x}-\bm{x}^k)
\end{multline}
where $\mathbf{H}_C$ is the Hessian matrix of the objective function $C(\bm{x})$:
\begin{equation}
    (\mathbf{H}_C)_{i,j}= \frac{\partial ^2 C}{\partial x_i x_j}
\end{equation}
Similarly, the constraints $g_j(\bm{x})$ and $h_j(\bm{x})$ can also be approximated by first order Taylor expansions:
\begin{subequations}\label{eq:taylor_cons}
\begin{alignat}{2}
h_j(\bm{x})& \approx h_j(\bm{x}^k) + \nabla h_j(\bm{x}^k)^{\mathrm{T}}(\bm{x}-\bm{x}^k)        \\
g_j(\bm{x}) & \approx g_j(\bm{x}^k) + \nabla g_j(\bm{x}^k)^{\mathrm{T}}(\bm{x}-\bm{x}^k) 
\end{alignat}
\end{subequations}
With the approximations made at the $k$-th iteration and let $\bm{d}^k = \bm{x}-\bm{x}^k$, a quadratic optimization subproblem (Equation \ref{eq:quad_sub}) can be formulated to find the search direction $\bm{d}^k$:
\begin{subequations}\label{eq:quad_sub}
\begin{alignat}{2}
&\underset{\bm{d}^k}{\text{minimize}} \quad C(\bm{x}^k) + \nabla C(\bm{x}^k)^{\mathrm{T}}\bm{d}^k + \frac{1}{2}(\bm{d}^k)^{\mathrm{T}}\mathbf{H}_{C(\bm{x}^k)}\bm{d}^k   \\
&\text{subject to: } \quad h_j(\bm{x}^k) + \nabla h_j(\bm{x}^k)^{\mathrm{T}}\bm{d}^k \leq 0   \\
& g_j(\bm{x}^k) + \nabla g_j(\bm{x}^k)^{\mathrm{T}}\bm{d}^k = 0
\end{alignat}
\end{subequations}
The quadratic programming subproblem (Equation \ref{eq:quad_sub}) is an approximation of the original optimization problem (Equation \ref{eq:nlp}) at iterate $\bm{x}^k$ by using the approximated forms of the objective function and the constraints. To further relax the problem for mathematical consideration, the Lagrangian of the original problem (Equation \ref{eq:nlp}) is considered:
\begin{equation}\label{eq:lagrangian}
    \mathcal{L}(\bm{x},\bm{\lambda}, \bm{\mu}) = C(\bm{x}) + \sum_{j=1}^{n_{ie}}\lambda_j h_j(\bm{x}) +\sum_{j={1}}^{n_e}\mu_j g_{j}(\bm{x})
\end{equation}
where $\bm{\lambda} \in \mathbb{R}^{n_{ie}}_{+}$ and $\bm{\mu}\in \mathbb{R}^{n_e}$  are the Lagrangian multiplier vectors for inequality constraints and equality constraints, respectively. Note that if the constraints in Equation \ref{eq:nlp} are satisfied, $\mathcal{L}(\bm{x},\bm{\lambda}, \bm{\mu}) \leq C(\bm{x})$, which implies by minimizing $\mathcal{L}(\bm{x},\bm{\lambda}, \bm{\mu})$, one can find a lower bound for the original objective function $C(\bm{x})$. The updated quadratic programming subproblem is formulated as follows by replacing $C(\bm{x})$ with $\mathcal{L}(\bm{x},\bm{\lambda},\bm{\mu})$:
\begin{multline}\label{eq:quad_sub_lag}
    \underset{\bm{d}^k}{\text{minimize}} \quad \mathcal{L}(\bm{x}^k,\bm{\lambda}^k, \bm{\mu}^k) + \nabla \mathcal{L}(\bm{x}^k,\bm{\lambda}^k, \bm{\mu}^k)^{\mathrm{T}}\bm{d}^k +\\
    \frac{1}{2}(\bm{d}^k)^{\mathrm{T}}\mathbf{H}_{\mathcal{L}(\bm{x}^k,\bm{\lambda}^k, \bm{\mu}^k)}\bm{d}^k   
\end{multline}
\begin{subequations}
\begin{align}
&\text{subject to: } \quad h_j(\bm{x}^k) + \nabla h_j(\bm{x}^k)^{\mathrm{T}}\bm{d}^k \leq 0   \\
& g_j(\bm{x}^k) + \nabla g_j(\bm{x}^k)^{\mathrm{T}}\bm{d}^k = 0
\end{align}
\end{subequations}
At an iterate $\bm{x}^k$, the $\mathcal{L}(\bm{x}^k,\bm{\lambda}^k, \bm{\mu}^k)$ term does not change so it can be eliminated from the quadratic programming subproblem. In addition, the Hessian is usually approximated in SQP methods and let $B_k \approx \mathbf{H}_{\mathcal{L}(\bm{x}^k,\bm{\lambda}^k, \bm{\mu}^k)}$, the subproblem is transformed into the following form:
\begin{subequations}\label{eq:quad_sub_lag_simple}
\begin{alignat}{2}
&\underset{\bm{d}^k}{\text{minimize}} \quad  \nabla \mathcal{L}(\bm{x}^k,\bm{\lambda}^k, \bm{\mu}^k)^{\mathrm{T}}\bm{d}^k + \frac{1}{2}(\bm{d}^k)^{\mathrm{T}}B_k\bm{d}^k    \\
&\text{subject to: } \quad h_j(\bm{x}^k) + \nabla h_j(\bm{x}^k)^{\mathrm{T}}\bm{d}^k \leq 0   \\
& g_j(\bm{x}^k) + \nabla g_j(\bm{x}^k)^{\mathrm{T}}\bm{d}^k = 0
\end{alignat}
\end{subequations}
At this iterate $\bm{x}^k$, the search direction $\bm{d}^k$ is then obtained by solving the approximation of the original problem in a quadratic programming form to find the next iterate $\bm{x}^{k+1}$. By iteratively solving the approximations of the original problems, the final solution will converge to the solution of the original problem \citep{boggs1995sequential}. The reason for formulating a subproblem as quadratitc programming is because quadratic programming is easy to solve due to the existence of rapid and accurate algorithms for this type of problems \citep{boggs1995sequential}. Usually, the approximated Hessian $B_{k+1}$ is updated at each iteration by the following principle that is essentially a secant approximation of the Hessian of the Lagrangian:
\begin{multline}\label{eq:Bk_update}
    B_{k+1}\bm{d}^k \sim \nabla \mathcal{L}(\bm{x}^{k+1},\bm{\lambda}^{k+1},\bm{\mu}^{k+1}) - \\
    \nabla \mathcal{L}(\bm{x}^{k},\bm{\lambda}^{k+1},\bm{\mu}^{k+1}) 
\end{multline}
In practice, there are different methods to iteratively update $B_{k+1}$ from $B_{k}$, such as the Broyden-Fletcher-Goldfarb-Shanno (BFGS) update, which is used by SLSQP. In our work, SLSQP \citep{kraft1988software} is selected as the optimizer for constrained shape optimization problems. The inputs of SLSQP are the gradients of the objective functions and the constraints: $\nabla C(\bm{x}^k)$, $\nabla h_{j}(\bm{x}^k)_{j \in\{1,...,n_{ie}\}}$, and $\nabla g_{j}(\bm{x}^k)_{j\in\{1,...,n_{e}\}}$. The output of SLSQP is the search direction $\bm{d}^k$ and the updated design variable $\bm{x}^{k+1}$. As a result, the key to obtain the optimized shape for structures is calculating the required gradient terms in SLSQP.
\subsection{Derivative calculation with AD}
Calculating the derivatives is of great significance for the implementation of gradient-based optimization algorithms. Traditionally, there are three methods for derivative evaluation \citep{baydin2018automatic}: i) manual derivation: derive the formulation of the derivatives by hand, which may be error-prone and time-consuming \citep{chandrasekhar2021auto}; ii) numerical differentiation: implement numerical algorithms such as finite difference to approximate the derivative values, which may be inaccurate due to truncation and round-off errors; numerical differentiation also does not scale well with the dimensions of the problems so it can be time-consuming: the work by \cite{pastrana_constrained_2022} shows that finite difference is slow for constrained form finding problems, especially when the number of design variables increases; iii) symbolic differentiation: use expression manipulation to derive the derivative expressions, but the expression may be complex and the problems of ``expression swell" may emerge \citep{baydin2018automatic,oberbichler2021efficient}. There exists a fourth method, automatic differentiation (AD), that can resolve the aforementioned problems, enabling fast and accurate derivative calculation.\par
\begin{figure}[h]
\centering
\includegraphics[width=1\linewidth]{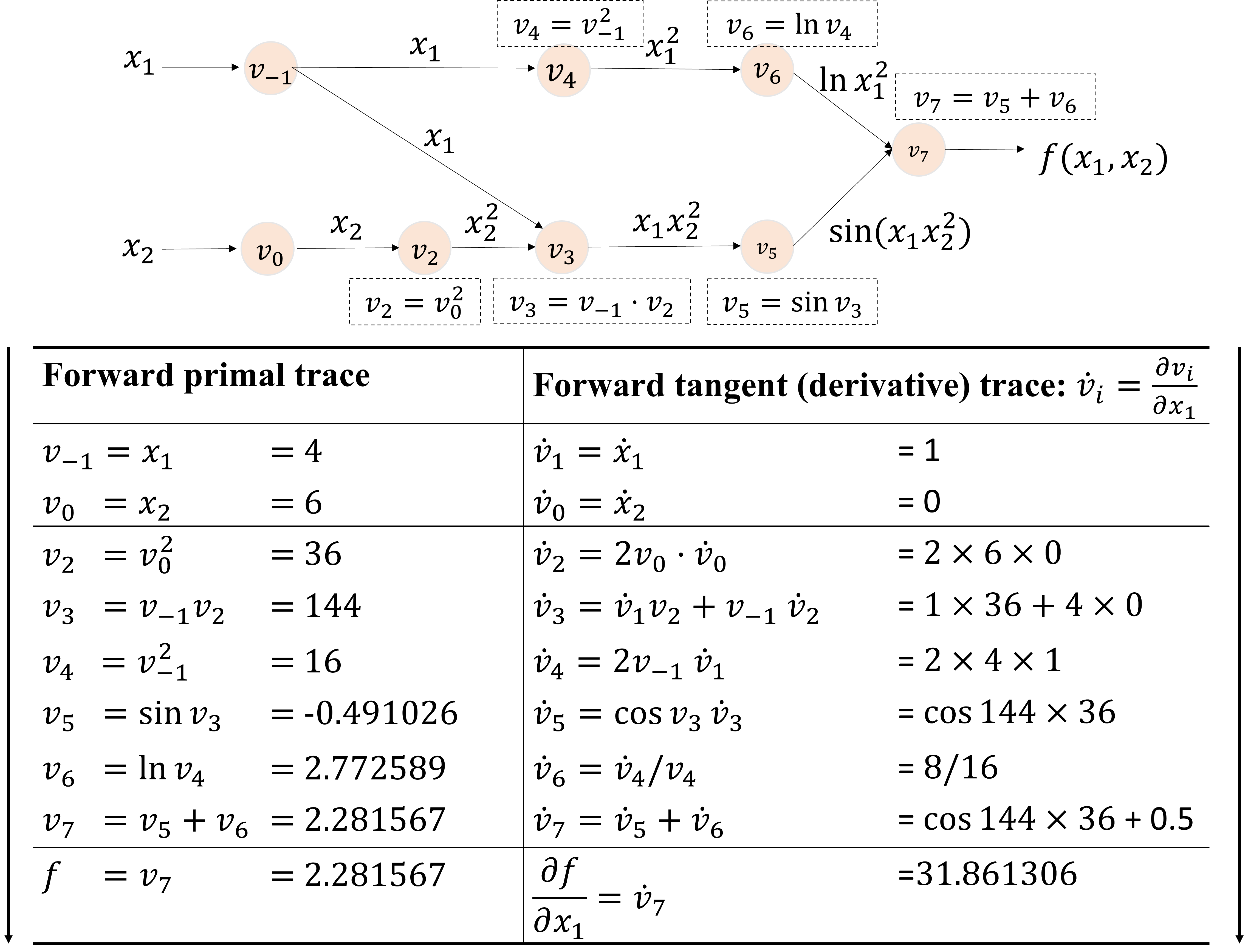}
\caption{Computational graph for $f(x_1,x_2)=\sin(x_1x_2^2) + \ln x_1^2$ and the table of forward trace for $\frac{\partial f}{\partial x_1}$ ; $v_{-1} ... v_{7}$ represent intermediate variables based on elementary operations (operation in dashed box in the graph)}
\label{fig:ad_graph}
\end{figure}
AD is an efficient tool for derivative computation of functions expressed as computer programs. AD exploits the fact that any derivative calculation can be decomposed as a set of basic derivative calculations so that one can trace these basic calculations sequentially to get the overall derivative by applying the chain rule \citep{baydin2018automatic}. AD gives the exact derivative value to the working precision of the computer so it is not numerical differentiation. It is also different from manual derivation or symbolic differentiation because it does not output a mathematical expression of the derivative. \par 
Figure \ref{fig:ad_graph} presents the derivative calculation of $f(x_1,x_2)=\sin(x_1x_2^2) + \ln x_1^2$ with respect to $x_1$ as an example to illustrate the key idea of AD. The function $f$ is decomposed into intermediate variables $v_{i \mid i\in\{-1,0,...,7\}}$, starting with $x_1$ and $x_2$, the inputs of the function $f$. Other intermediate variables are the outputs of elementary operations (marked in the dashed boxes) such as addition, multiplication and trigonometric functions whose derivatives are already known. The inputs of these elementary operations are the intermediate variables in the previous step. One can trace these intermediate variables and their derivatives in a sequential order and apply the chain rule to get the overall derivatives. This process is often referred as the \textit{evaluation trace} and the table in Figure \ref{fig:ad_graph} illustrates the evaluation trace process in the forward mode. The \textit{forward primal trace} calculates the values of intermediate variables themselves whereas the \textit{forward tangent trace} evaluates the derivatives of intermediate variables. At the end, the overall derivative $\frac{\partial f}{\partial x_1}$ is obtained from the evaluation trace. In this example, we present the ``manual" evaluation trace. In practice, the evaluation traces are conducted by computer programs so the gradients needed by optimizers can be computed efficiently. \par
\section{The proposed framework} \label{sec:jax-sso}
\subsection{Introduction to JAX}\label{subsec:jax}
JAX \citep{jax2018github}, developed by Google, is a Python library that aims at providing AD and high-performance computing to the research community. JAX can automatically differentiate Python and NumPy \citep{harris2020array} functions, making it an ideal candidate for calculating derivatives in structural shape optimization. In addition, JAX supports just-in-time (JIT) compilation \citep{lam2015numba} via accelerated linear algebra (XLA), a compiler originally used in TensorFlow \citep{Abadi_TensorFlow_Large-scale_machine_2015}, to improve the speed and memory usage of computer programs. The AD and XLA features make JAX a good candidate for gradient evaluation in gradient-based structural shape optimization. \par
\begin{table*}[h]
\centering 
\caption{Key methods in JAX}\label{table:jax_methods}

\begin{tabular}{{l}{l}}
\hline
JAX methods              & Description \\
\hline
{\ttfamily jax.numpy} & Array programming package in JAX, its syntax is similar to {\ttfamily numpy}\\
{\ttfamily jax.jit/@jit} & JIT feature that compiles operations in Python functions to\\
& fast machine codes with XLA, to speed up the operations\\
{\ttfamily jax.vmap} & Method to vectorize the operations for faster implementation\\
{\ttfamily jax.jacfwd} & Method to calculate the Jacobian\\
{\ttfamily @register\_pytree\_node\_class} & Register a new class as a valid type for JAX \\
\hline
\end{tabular}
\end{table*}
We first introduce some key methods in JAX (Table \ref{table:jax_methods}). In array programming \citep{harris2020array}, {\ttfamily numpy} has been widely used. JAX has a similar array programming package called {\ttfamily jax.numpy}, which can be used to store values such as structural stiffness matrix using {\ttfamily jax.numpy.array}. The main difference between {\ttfamily jax.numpy} and {\ttfamily numpy} is that values and operations in {\ttfamily jax.numpy} can be \textit{traceable}, assisting the implementation of AD using the \textit{chain rule}. However, the traceable feature of {\ttfamily jax.numpy} also means extra operations and memory are needed, making the computational speed a significant issue. In response, {\ttfamily jax.jit} or {\ttfamily @jit} can be used to compile a sequence of Python operations into optimized machine codes that operate faster once they have been compiled \citep{chandrasekhar2021auto}. The first call of ``jitted" function is the compilation process, which is relatively slow, but once it has been compiled, the future calls will be extremely fast. Another way to speed up the operations is using {\ttfamily jax.vmap} to vectorize the operations. Instead of operating on individual elements sequentially, vectorization enables operations on the entire array simultaneously \citep{harris2020array} to improve the speed. To calculate the gradient of functions, {\ttfamily jax.jacfwd} can be used. In order to implement the methods like {\ttfamily jax.vmap} for Python objects of any customized Python class, {\ttfamily @register\_pytree\_node\_class} decorator can be used to register the customized Python class as a valid JAX type. In addition, {\ttfamily tree\_flatten} and {\ttfamily tree\_unflatten} methods are needed to tell JAX how to flatten and unflatten the attributes of customized objects. A toy-code example is attached in Appendix \ref{app:app_1} to illustrate how the JAX methods work by presenting a simple problem: calculating
how the length of a 2D line changes with the nodal coordinates.
\subsection{The framework and the JAX-SSO package}
The proposed framework to assist gradient-based structural shape optimization problems is illustrated in Algorithm \ref{algo:JAX-SSO_Procedure}. The source code of this framework is available at: \href{https://github.com/GaoyuanWu/JaxSSO}{https://github.com/GaoyuanWu/JaxSSO}. A JAX-supported Python package is developed for calculating the derivatives of the global stiffness matrix with respect to the nodal coordinates, which is named as JAX-SSO.\par 
The iterations of the structural shape optimization start after initializing the design variables $\bm{x}^0$, the optimizer and the convergence criteria ($N_{max}$, $\epsilon_C$). Firstly, a sensitivity analysis model is built via JAX-SSO and the sensitivity of the global stiffness matrix $\frac{\partial \bm{K}}{\partial \bm{x}}$ can be obtained using AD. Secondly, a finite element model is built and solved for the displacement vector $\bm{u}$. The sensitivity of the objective function can then be calculated according to Equation \ref{eq:dcdxi_simplified} by combining $\frac{\partial \bm{K}}{\partial \bm{x}}$ and $\bm{u}$, using the derived sensitivity from the adjoint method. If constraints are specified, the next step is to calculate the gradient of the constraints using AD, if required by the optimization algorithm selected. Lastly, the design variables at step $k$, $\bm{x}^k$, are updated by the optimizer to $\bm{x}^{k+1}$. The compliance of step $k$ is stored and the next iteration starts until the convergence criteria is met.\par
\begin{figure*}[h]
\centering
\includegraphics[width=0.85\linewidth]{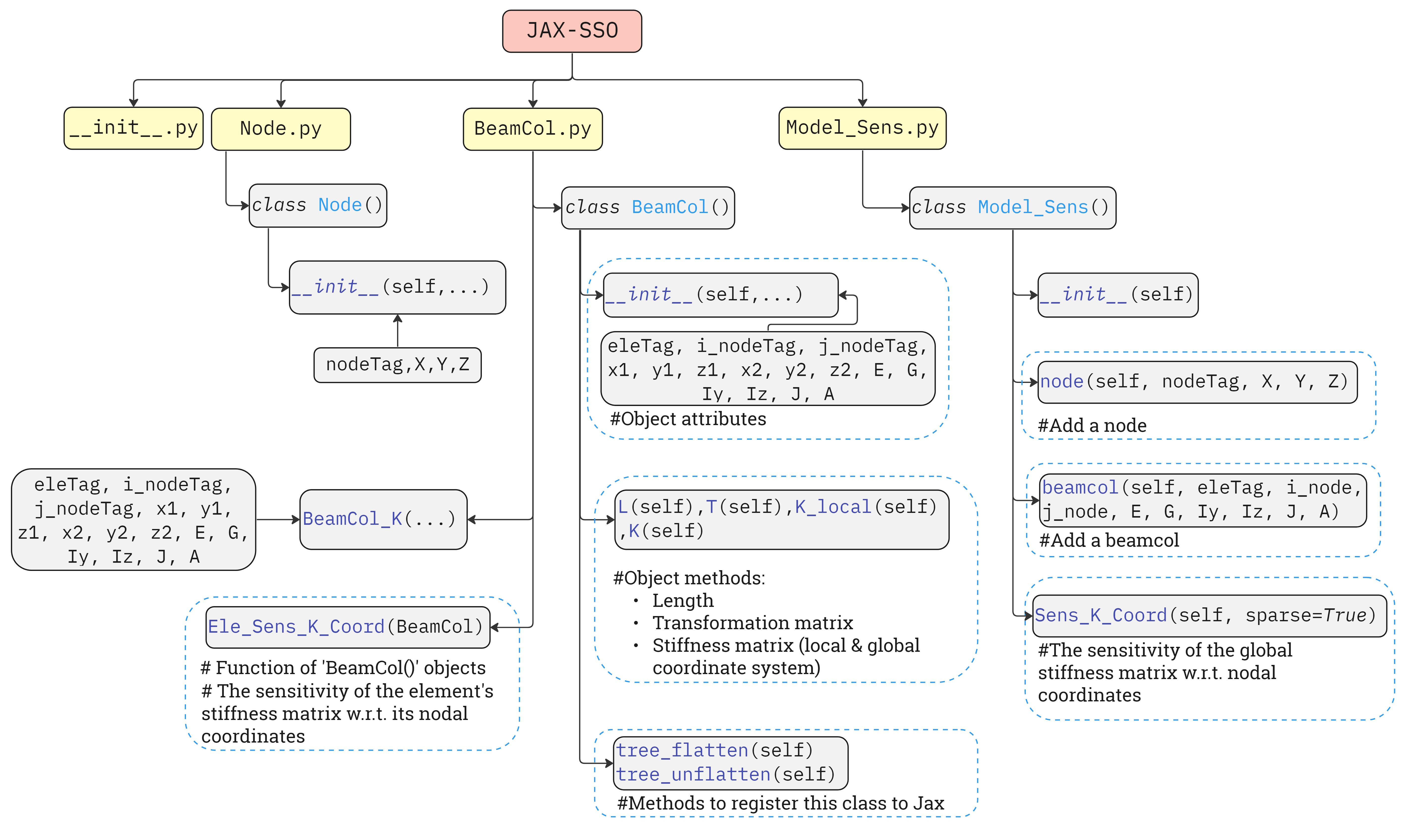}
\caption{The structure of JAX-SSO Python package}
\label{fig:jax_sso_code_structure}
\end{figure*}

 
\begin{algorithm}
\caption{The proposed framework for structural shape optimization}\label{algo:JAX-SSO_Procedure}
\begin{algorithmic}
\State Initialize $N_{max}$, $\epsilon_C$, $\bm{x}^0$, Optimizer
\State $k\gets0$
\While{$ k \leq N_{max}$ \& $ \mid C^k-C^{k-1}\mid \geq \epsilon_C$} 
\State Sensitivity model (JAX-SSO) \hfill $\triangleright$ See Code \ref{code:addmodel}
\State Obtain $\frac{\partial \bm{K}}{\partial \bm{x}^k}$ (JAX-SSO)\hfill $\triangleright$ See Code \ref{code:senskcoord_model}
\State Build FEA model
\State Obtain $\bm{u}^k$ from FEA
\State Obtain $\nabla C(\bm{x}^k)$ using the adjoint method \hfill $\triangleright$ See Equation \ref{eq:dcdxi_simplified}
\State Calculate $\nabla h_{j}(\bm{x}^k)_{\mid j\in\{1,...,n_{ie}\}}$, if needed
\State Calculate $\nabla g_{j}(\bm{x}^k)_{\mid j\in\{1,...,n_{e}\}}$, if needed
\State $\bm{x}^{k+1} \gets$ \text{Optimizer}($C(\bm{x}^k)$, $h_j(\bm{x}^k)$, $\nabla C(\bm{x}^k)$, $\nabla h_{j}(\bm{x}^k))$
\State $C^k \gets \frac{1}{2}\bm{f}^\mathrm{T}\bm{u}^k$
\State $k \gets k+1$
\EndWhile
\end{algorithmic}
\end{algorithm}
We then introduce the JAX-SSO package, which is of great significance in the proposed framework as it outputs $\frac{\partial \bm{K}}{\partial \bm{x}}$ to assist gradient-based shape optimization. The code structure of JAX-SSO is illustrated in Figure \ref{fig:jax_sso_code_structure}, which is similar to FEA solvers: it has different classes representing the nodes, the elements and the model. Here we use the beam-column as an example of structural elements but the proposed framework can be extended to structures with any structural elements, such as springs, plates, shells and solids.\par
In {\ttfamily Node.py}, there is a class called {\ttfamily Node()} for structural nodes, in which each node is defined by a unique tag and nodal coordinates. \par
\begin{figure*}
\centering
\includegraphics[width=0.85\linewidth]{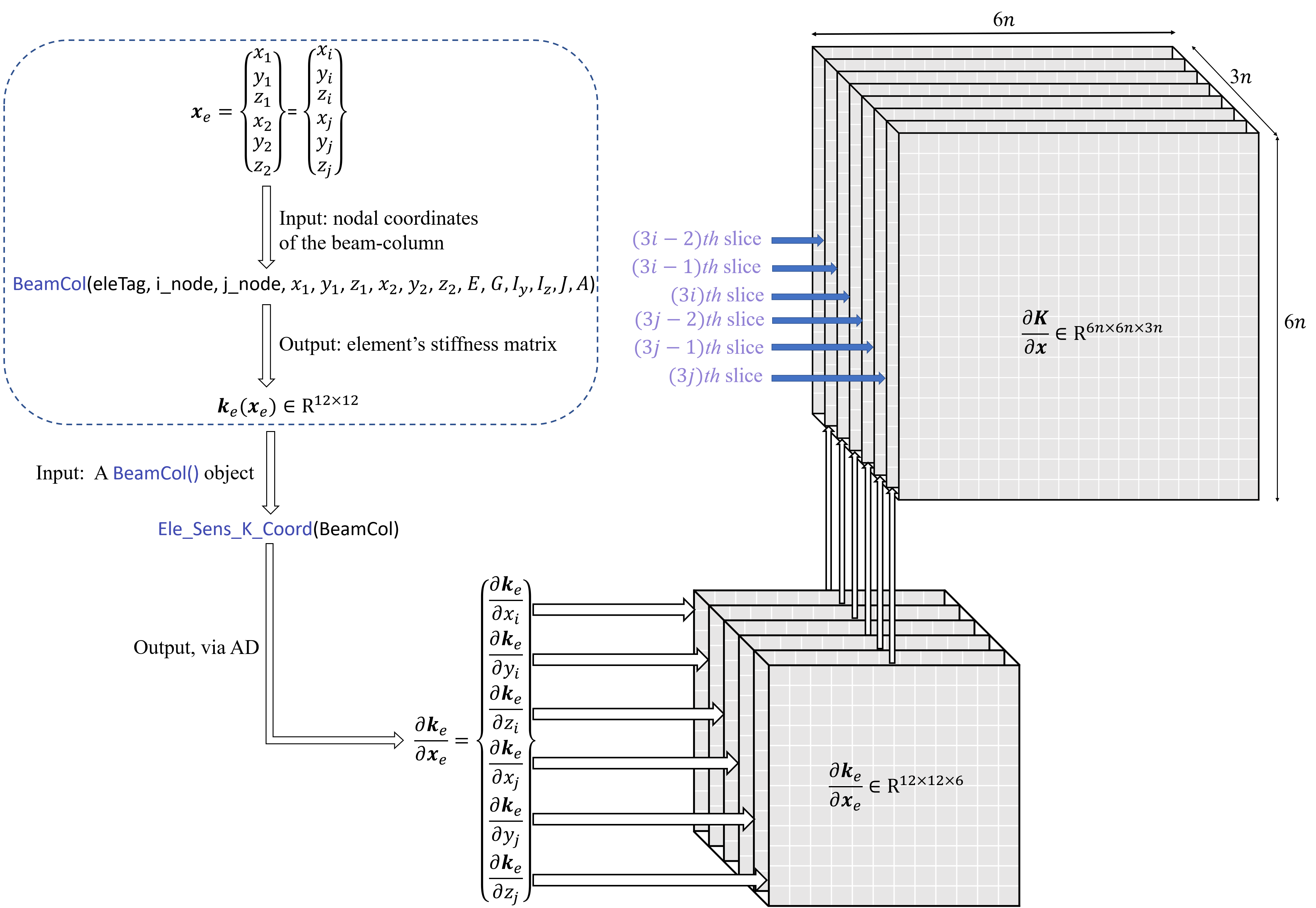}
\caption{The assembly of $\frac{\partial \bm{K}}{\partial \bm{x}}$ from $\frac{\partial \bm{k}_e}{\partial \bm{x}_e}$}
\label{fig:dkdx}
\end{figure*}

 
The {\ttfamily BeamCol.py} module consists of a class called {\ttfamily BeamCol()} for beam-column elements and functions related to the element, similar to the structure of Code \ref{code:toy_jax} in Appendix \ref{app:app_1} where we define a class for a line and functions calculating the derivatives for lines. To create a {\ttfamily BeamCol()} object, attributes like the tag of this element, the cross-sectional properties, and the connectivity are needed. Methods for a {\ttfamily BeamCol()} object include the length of the element, the coordinate transformation matrix, the stiffness matrix in local coordinate system and the stiffness matrix in global coordinate system. To conduct gradient-based structural shape optimization, we need to calculate $\frac{\partial \bm{K}}{\partial \bm{x}}$. Assume the design variables are all the nodal coordinates in the structure, we then have $\bm{x} \in \mathbb{R}^{3n}$ and $\frac{\partial \bm{K}}{\partial \bm{x}}\in \mathbb{R}^{6n\times6n\times3n}$. Figure \ref{fig:dkdx} illustrates how the entries of $\frac{\partial \bm{K}}{\partial \bm{x}}$ are formed and the idea behind is that the global stiffness matrix $\bm{K}$ is essentially an assembly of the local stiffness matrix $\bm{k}_e$ of all the elements in the structural system. Let us consider a beam-column that consists of two nodes and their indices are $i$ and $j$. With other properties of this element unchanged, the local stiffness matrix $\bm{k}_e \in \mathbb{R}^{12\times12}$ of this element is a function of the vector storing the nodal coordinates of this element, $\bm{x}_e \in \mathbb{R}^{6}$. In {\ttfamily BeamCol.py}, a function called {\ttfamily Ele\_Sens\_K\_Coord} is then defined for {\ttfamily BeamCol()} objects (Code \ref{code:ele_sens}), returning the sensitivity of element's local stiffness matrix with respect to the element's coordinates, $\frac{\partial \bm{k}_e}{\partial \bm{x}_e} \in \mathbb{R}^{12\times12\times6}$, by using AD. Since the element's influence on the global stiffness matrix $\bm{K}$ is determined by its two nodes (node $i$ and node $j$), we only need to look at six sub-matrices (slices) of $\frac{\partial \bm{K}}{\partial \bm{x}}$ that represent how the global stiffness matrix is influenced by the coordinates of node $i$ and node $j$: the $(3i-2)$\textit{th}, $(3i-1)$\textit{th} and $3i$\textit{th} sub-matrices of $\frac{\partial \bm{K}}{\partial \bm{x}}$ represent the influence of the X, Y and Z coordinate of node $i$ on the global stiffness matrix $\bm{K}$, respectively; similarly, the $(3j-2)$\textit{th}, $(3j-1)$\textit{th} and $3j$\textit{th} sub-matrices of $\frac{\partial \bm{K}}{\partial \bm{x}}$ represent the influence of the X, Y and Z coordinate of node $j$ on the global stiffness matrix $\bm{K}$, respectively. Each sub-matrix of $\frac{\partial \bm{K}}{\partial \bm{x}}$ has a size of $6n$ by $6n$ and corresponds to one sub-matrix of $\frac{\partial \bm{k}_e}{\partial \bm{x}_e}$ as shown in Figure \ref{fig:dkdx}. Based on the indices of the nodes that the element is made of, the entries in $\frac{\partial \bm{k}_e}{\partial \bm{x}_e}$ are then added into the global stiffness sensitivity $\frac{\partial \bm{K}}{\partial \bm{x}}$. Equation \ref{eq:assign_dk_dx_1}-\ref{eq:assign_dk_dx_4} illustrates how to add the entries from $\frac{\partial \bm{k}_e}{\partial x_i}$ to $\frac{\partial \bm{K}}{\partial \bm{x}}$ as an example of the assembly of $\frac{\partial \bm{K}}{\partial \bm{x}}$:
\begin{multline}\label{eq:assign_dk_dx_1}
    (\frac{\partial \bm{K}}{\partial \bm{x}})_{6(i-1)+m,6(i-1)+n,3i-2} \gets \\    (\frac{\partial \bm{K}}{\partial \bm{x}})_{6(i-1)+m,6(i-1)+n,3i-2} + (\frac{\partial \bm{k}_e}{\partial x_i})_{m,n} \\ \text{when } m\leq6, n\leq6
\end{multline}
\begin{multline}\label{eq:assign_dk_dx_2}
     (\frac{\partial \bm{K}}{\partial \bm{x}})_{6(i-1)+m,6(j-1)+n-6,3i-2} \gets \\    (\frac{\partial \bm{K}}{\partial \bm{x}})_{6(i-1)+m,6(j-1)+n-6,3i-2} + (\frac{\partial \bm{k}_e}{\partial x_i})_{m,n} \\ \text{when }m\leq6, n>6
\end{multline}
\begin{multline}\label{eq:assign_dk_dx_3}
    (\frac{\partial \bm{K}}{\partial \bm{x}})_{6(j-1)+m-6,6(j-1)+n-6,3i-2} \gets \\    (\frac{\partial \bm{K}}{\partial \bm{x}})_{6(j-1)+m-6,6(j-1)+n-6,3i-2} + (\frac{\partial \bm{k}_e}{\partial x_i})_{m,n} \\ \text{when }m>6, n>6
\end{multline}
\begin{multline}\label{eq:assign_dk_dx_4}
    (\frac{\partial \bm{K}}{\partial \bm{x}})_{6(j-1)+m-6,6i+n,3i-2} \gets \\    (\frac{\partial \bm{K}}{\partial \bm{x}})_{6(j-1)+m-6,6i+n,3i-2} + (\frac{\partial \bm{k}_e}{\partial x_i})_{m,n} \\ \text{when }m>6, n\leq6
\end{multline}
where $m\in \{1,2,...,12\}$ and $n \in \{1,2,...,12\}$ are the row and columns of the sub-matrix $\frac{\partial \bm{k}_e}{\partial x_i}$.
If the assembly process shown above is repeated for each element $e$ in the system, $\frac{\partial \bm{K}}{\partial \bm{x}}$ can be obtained.\par
The {\ttfamily Model\_Sens.py} module is the main module of the JAX-SSO package that directly interacts with the users. A sensitivity model can be built by initializing an object from class {\ttfamily Model\_Sens()} (Code \ref{code:addmodel}). Nodes and structural elements can then be added to the sensitivity model, as shown in Code \ref{code:addmodel}. After having added all the structural elements to the model, one can then calculate the sensitivity terms needed in the optimization process. From the user's side, a one-liner code illustrated in Code \ref{code:senskcoord_model} helps them to obtain the sensitivity of the global stiffness matrix with respect to all the nodal coordinates in the system, $\frac{\partial \bm{K}}{\partial \bm{x}} \in \mathbb{R}^{6n\times 6n \times 3n}$, which is usually stored by a sparse matrix.\par
\begin{lstlisting}[language=Python, label={code:ele_sens}, caption=Functions that calculate the sensitivity of element's stiffness matrix with respect to its nodal coordinates]
#Stiffness matrix of the element
def BeamCol_K(eleTag, i_nodeTag, j_nodeTag, x1, y1, z1, x2, y2, z2,E,G,Iy,Iz,J,A):

    #Create a beam-column
    this_beamcol = BeamCol(eleTag, i_nodeTag, j_nodeTag, x1, y1, z1, x2, y2, z2, E, G, Iy, Iz, J, A)

    #Get the local stiffness matrix of this element
    return this_beamcol.K()

def Ele_Sens_K_Coord(BeamCol):
    '''
    Return the sensitivity of element's 
    local stiffness matrix (in global coordinates) 
    w.r.t. the element's coordinates.
   
    Inputs:
        BeamCol: A 'BeamCol' object
    '''
    
    #Properties of this beam column
    eleTag = BeamCol.eleTag
    i_nodeTag, j_nodeTag = [BeamCol.i_nodeTag,BeamCol.j_nodeTag]
    x1,y1,z1,x2,y2,z2 = [BeamCol.x1,BeamCol.y1,BeamCol.z1,BeamCol.x2,BeamCol.y2,BeamCol.z2]
    E = BeamCol.E
    G = BeamCol.G
    Iy = BeamCol.Iy 
    Iz = BeamCol.Iz 
    J = BeamCol.J
    A = BeamCol.A
    
    #Calculate the sensitivity
    #argnums indicates the variables to which the Jacobian is calculated
    #and in our case they are (x1,...,z2)
    return jacfwd(BeamCol_K,argnums=(3,4,5,6,7,8))(eleTag, i_nodeTag, j_nodeTag, x1, y1, z1, x2, y2, z2, E, G, Iy, Iz, J, A)
\end{lstlisting}
\begin{lstlisting}[language=Python, label={code:addmodel}, caption={Creating a sensitivity model, adding nodes and elements in JAX-SSO (user's side)}]
import JaxSSO as sso #JAX-SSO module

sens_model = sso.Model_Sens.Model_Sens() #A JAX-SSO model
sens_model.node(nodeTag,x,y,z) #adding a node
sens_model.beamcol(i,i_node,j_node,E,G,Iy,Iz,J,A) #adding a beam-colomn element

\end{lstlisting}

\begin{lstlisting}[language=Python, label={code:senskcoord_model}, caption={Obtain the sensitivity of the global stiffness matrix with respect to nodal coordinates (user's side)}]
sens_K = sens_model.Sens_K_Coord(sparse=True)
\end{lstlisting}
\begin{figure*}[h]
\centering
\includegraphics[width=0.8\linewidth]{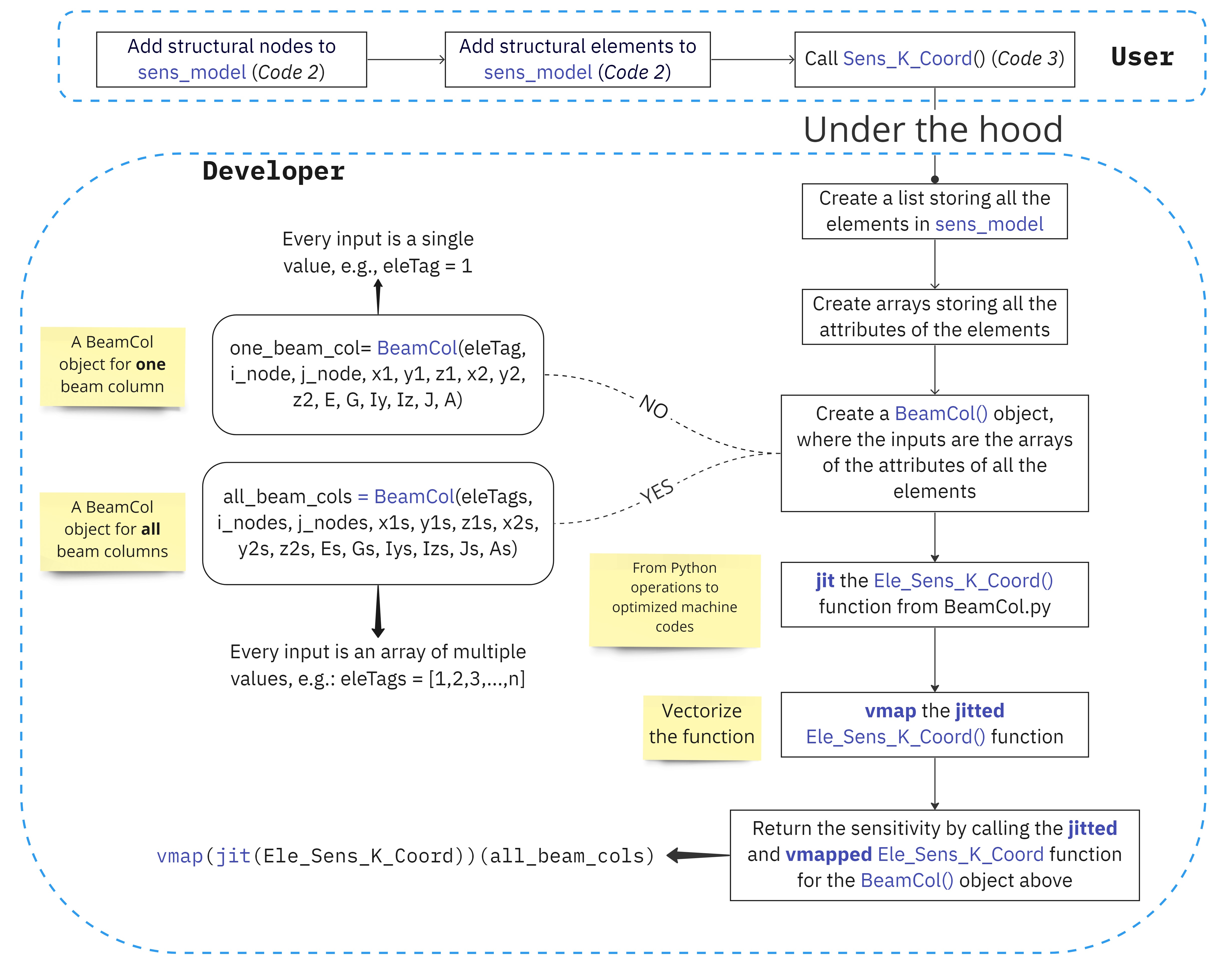}
\caption{The structure of {\ttfamily Sens\_K\_Coord}}
\label{fig:jit_vmap_sens}
\end{figure*}

Under the hood of {\ttfamily Sens\_K\_Coord} in Code \ref{code:senskcoord_model}, there is an efficient and compact structure for sensitivity calculation, which can be seen in Figure \ref{fig:jit_vmap_sens}. After having added all the structural elements in the sensitivity model, an array is created to store all the elements, which is further used to extract the attributes of all the elements in the structural system. Later, a new {\ttfamily BeamCol()} object that integrates all the beam-columns is created. For a single beam-column, the {\ttfamily Ele\_Sens\_K\_Coord} function is able to output the derivatives of the local stiffness matrix $\bm{k}_e$ with respect to this element's coordinates. To make this operation faster, we use {\ttfamily jit} to compile the Python operation into optimized binary machine codes. In addition, in order to apply {\ttfamily Ele\_Sens\_K\_Coord} to the integrated {\ttfamily BeamCol()} object so that the derivative calculation for every structural element can be conducted simultaneously, we implement {\ttfamily vmap} to vectorize the element-wise operation {\ttfamily Ele\_Sens\_K\_Coord}. This efficient and compact structure of {\ttfamily Sens\_K\_Coord} is of great significance in the proposed framework, as it realizes accurate and fast calculation of $\frac{\partial \bm{K}}{\partial \bm{x}}$ using AD, which is advantageous over manual derivation and numerical differentiation. The call of {\ttfamily Sens\_K\_Coord} can be further boosted by hardware-acceleration, such as GPU-acceleration enabled by NVidia GPUs. We will show how the proposed structure with {\ttfamily jit}, {\ttfamily vmap}, and GPU-acceleration improves the speed of derivative calculation in Section \ref{subsec:jax_perform}.\par
This framework is efficient and novel in a sense that it uses a combination of mathematical derivations via the adjoint method and JAX-boosted AD. Instead of calculating $\frac{\partial C}{\partial \bm{x}}$ directly using AD, we only use AD to calculate the derivatives related to the stiffness matrix ($\frac{\partial \bm{K}}{\partial \bm{x}}$) in Equation \ref{eq:dcdxi_simplified}. The advantages of doing so are as follows. Firstly, the derivative calculation of $\frac{\partial \bm{K}}{\partial \bm{x}}$ can be decomposed into efficient element-wise operations because the global stiffness matrix $\bm{K}$ is an assembly of local stiffness matrix ${\bm{k}_e}$ of each element $e$. These element-wise operations can then be optimized and boosted by vectorization, just-in-time compilation, and GPU acceleration. Secondly, the designer can couple the derivative-calculation package JAX-SSO with any FE solver one prefers to obtain the displacement $\bm{u}$ and can then calculate the sensitivity ($\frac{\partial C}{\partial \bm{x}}$) using Equation \ref{eq:dcdxi_simplified}. Moreover, compared to the direct method, the adjoint method reduces the complexity of the derivative calculation since it avoids solving the system of linear equations (Equation \ref{eq:kdudxi}). The proposed framework is also flexible because new element-types can be added to JAX-SSO to support the optimization of new structural systems without carrying out mathematical derivations as AD calculates the derivatives ``automatically".
\subsection{Performance of JAX-SSO}\label{subsec:jax_perform}
\begin{figure}[h]
\centering
\includegraphics[width=0.9\linewidth]{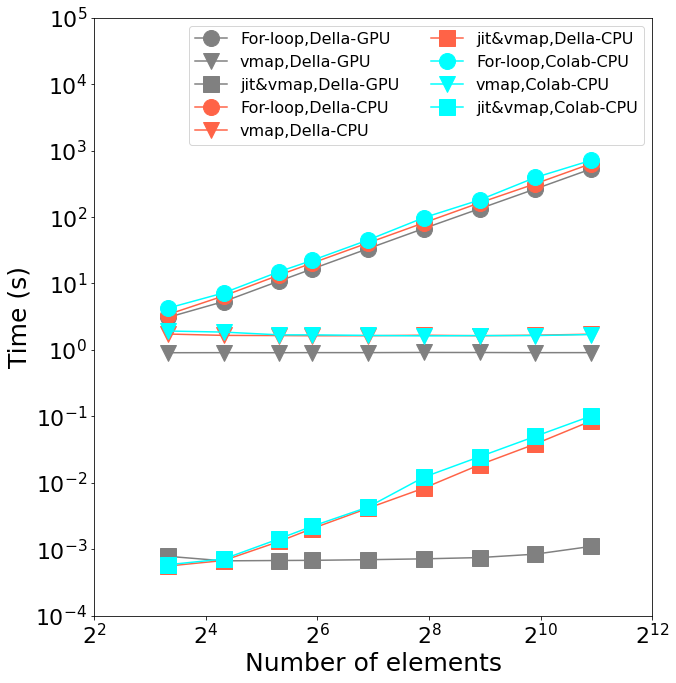}
\caption{Performance of the JAX-SSO module}
\label{fig:performance}
\end{figure}

We now illustrate the performance of JAX-SSO by presenting its speed of calculating $\frac{\partial \bm{K}}{\partial \bm{x}}$ (Figure \ref{fig:performance}).
Three different code structures are compared: i) using a for-loop to calculate the derivatives of the local stiffness matrix of every structural element sequentially, ii) using {\ttfamily vmap} to vectorize the operations so that the derivatives are calculated simultaneously, and iii) using {\ttfamily jit} to compile the vectorized codes into optimized operations using XLA, which is the structure adopted by JAX-SSO. Furthermore, we use three different devices to showcase the speed improved by GPU acceleration: i) CPU: 
2.25 GHz AMD EPYC 7B12 CPU on Google Colab, ii) CPU: 2.6 GHz AMD EPYC Rome CPU on Princeton University's Della cluster, and iii) GPU: Nvidia's A100 GPU on Princeton University's Della cluster.
As can be seen in Figure \ref{fig:performance}, the performance of the structure used by JAX-SSO (with {\ttfamily vmap} and {\ttfamily jit}) outperforms other structures significantly. Using for-loops with the CPU on Google Colab, the time needed for a structural system with 1920 structural elements is 711 seconds. With vectorization by {\ttfamily vmap}, the time is reduced to 1.7 seconds, which drops significantly but it is still not ideal. With the JAX-SSO structure, the computation only takes 0.101 seconds. The computation can be further accelerated by GPUs: the time needed for derivative calculation is $1.1 \times 10^{-3}$ seconds for a structural system with 1920 structural elements and the resulting $\frac{\partial \bm{K}}{\partial \bm{x}}$ has a dimension of $\mathbb{R}^{11526\times11526\times5763}$. Meanwhile, with the increase of the dimension of the problem, the time needed does not significantly increase if one operates on GPU because the computation capability of this GPU has not been fully exploited.
This performance assessment further consolidates the idea that the implementation of AD and XLA efficiently helps structural designers to calculate the derivatives that used to be time-consuming and difficult to calculate. The performance of JAX-SSO can then facilitate the employment of gradient-based shape optimization in structural and architectural design.
\section{Validation and examples}\label{sec:test}
In this section, we use several examples to validate and test the capability of the the proposed framework for structural shape optimization. The properties of the structural members used can be seen in Table \ref{table:members} unless otherwise specified. For constrained shape optimization, the SLSQP algorithm from the NLopt library \citep{johnson2014nlopt} is used. The open-source solver PyNite is used for FEA.
\begin{table}[h]
\centering 
\caption{Properties of the structural members}\label{table:members}
\begin{tabular}{{c}{c}{c}}
\hline
Property       & Value & Unit \\
\hline
$E$, Young's modulus& 37900 & Mpa\\
$G$, Shear modulus& 14577 & Mpa\\
$I_y$, Moment of inertia& 0.0072 & m$^4$\\
$I_z$, Moment of inertia& 0.0032 & m$^4$\\
$A$, Sectional area& 0.24 & m$^2$\\
\hline
\end{tabular}
\end{table}
\subsection{2D arch and barrel arch}
\begin{figure}[h]
     \centering
     \includegraphics[width=0.4\textwidth]{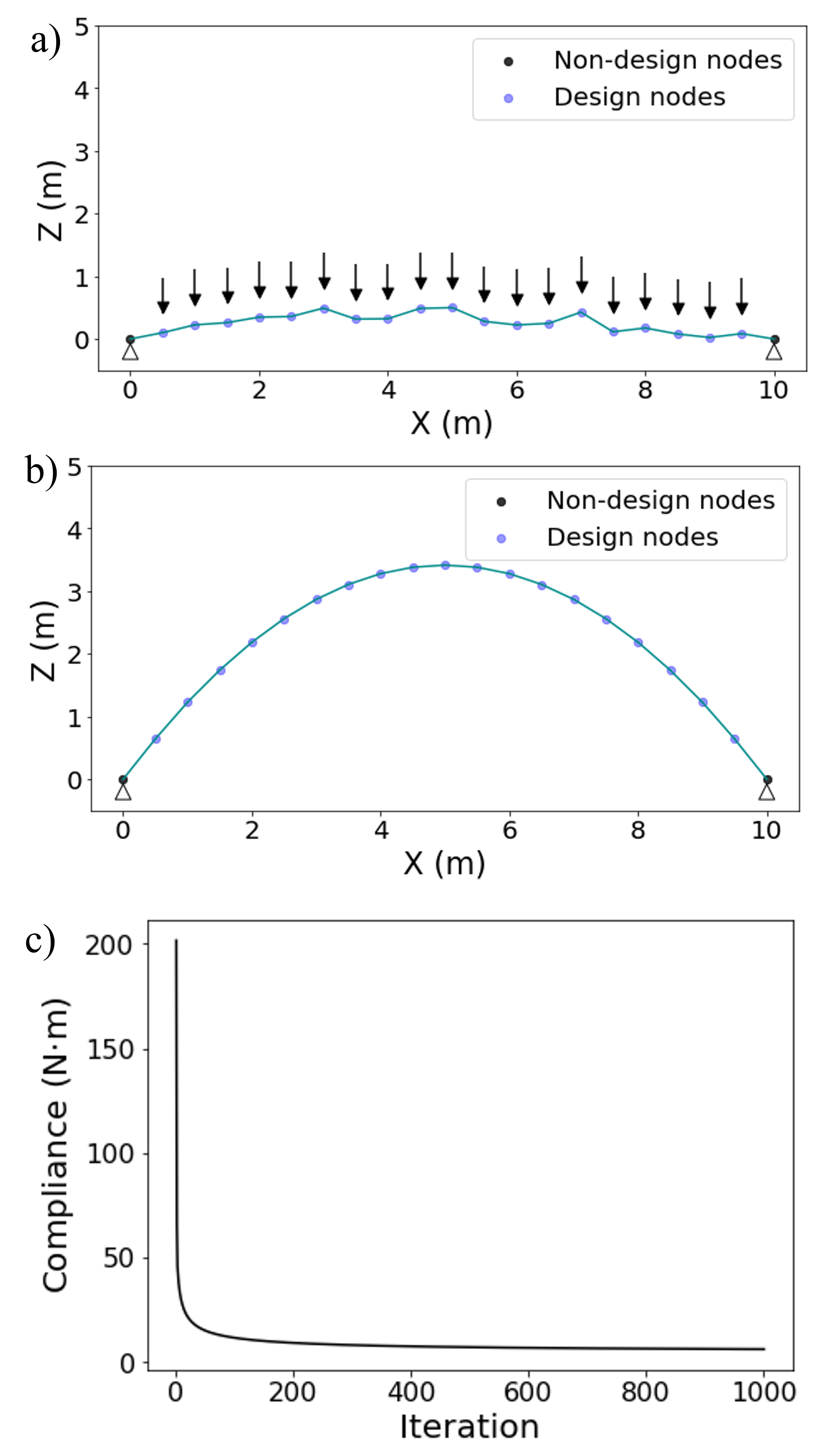}
    \caption{Form finding of 2D arch using gradient descent: a) initial structure; b) optimized structure; c) Optimization history of compliance}
    \label{fig:2darch}
\end{figure}
The proposed framework is first validated by finding the form of arches. The first example can be seen in Figure \ref{fig:2darch}, where the initial structure (Figure \ref{fig:2darch}a) is a simply-supported bridge spanning 10m with 21 nodes. The leftmost and rightmost nodes are pinned and the other nodes are the design nodes with randomly generated Z coordinates. A 10KN nodal load along -Z axis is applied to each design node. The design variables for this problem are the Z coordinates of the design nodes while the X coordinates remain unchanged. Gradient descent is used as the optimizer and after 1000 iterations, an arch bridge is found (Figure \ref{fig:2darch}b). The initial compliance of the bridge is 201 N$\cdot$m whereas the compliance of the optimized arch is only 6.32 N$\cdot$m, decreasing by 96.9\%. As has been known in engineering practice and confirmed by equilibrium-state based form finding methods, the optimal geometry for a two-end-pinned structure under gravity is an arch/catenary. The validity of the framework is then confirmed by this benchmark example. Unlike equilibrium-state based form finding methods, this framework does not impose the structural members to only resist membrane forces, it is the ``strain energy minimization" process naturally outputs the arch form that is membrane-force dominant.
\begin{figure}[h]
     \centering
     \includegraphics[width=0.5\textwidth]{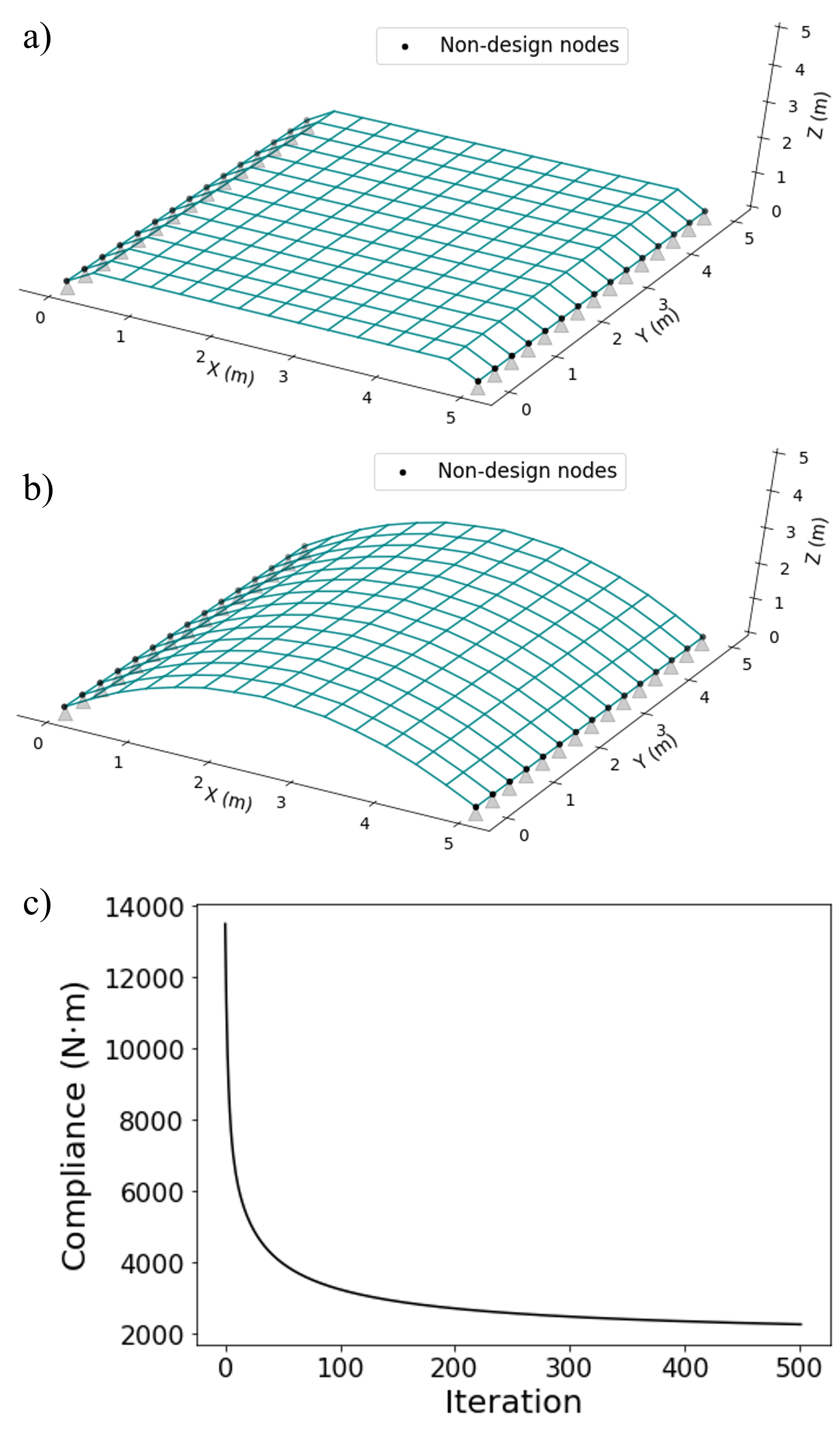}
    \caption{Form finding of 3D arch using gradient descent: a) initial structure; b) optimized structure; c) optimization history of compliance}
    \label{fig:3darch}
\end{figure}
The framework is further applied to the form finding of a 3D barrel arch (Figure \ref{fig:3darch}). The initial structure (Figure \ref{fig:3darch}a) spans 5 m in both the X and the Y direction and there are in total 225 structural nodes in the system. The nodal load is the same as the previous 2D example. Non-design nodes are pinned and the design nodes have the same randomly generated Z coordinates. The design variables are the Z coordinates of the design nodes. Similarly, gradient descent is used and the compliance is reduced from 13495 N$\cdot$m by 83.2\% to 2259 N$\cdot$m after 500 iterations (Figure \ref{fig:3darch}c). The optimized structure is in the form of a barrel arch, which can be seen in Figure \ref{fig:3darch}b.
\subsection{Free-form gridshell inspired by Mannheim Multihalle}
\begin{figure}[h]
     \centering
     \includegraphics[width=0.4\textwidth]{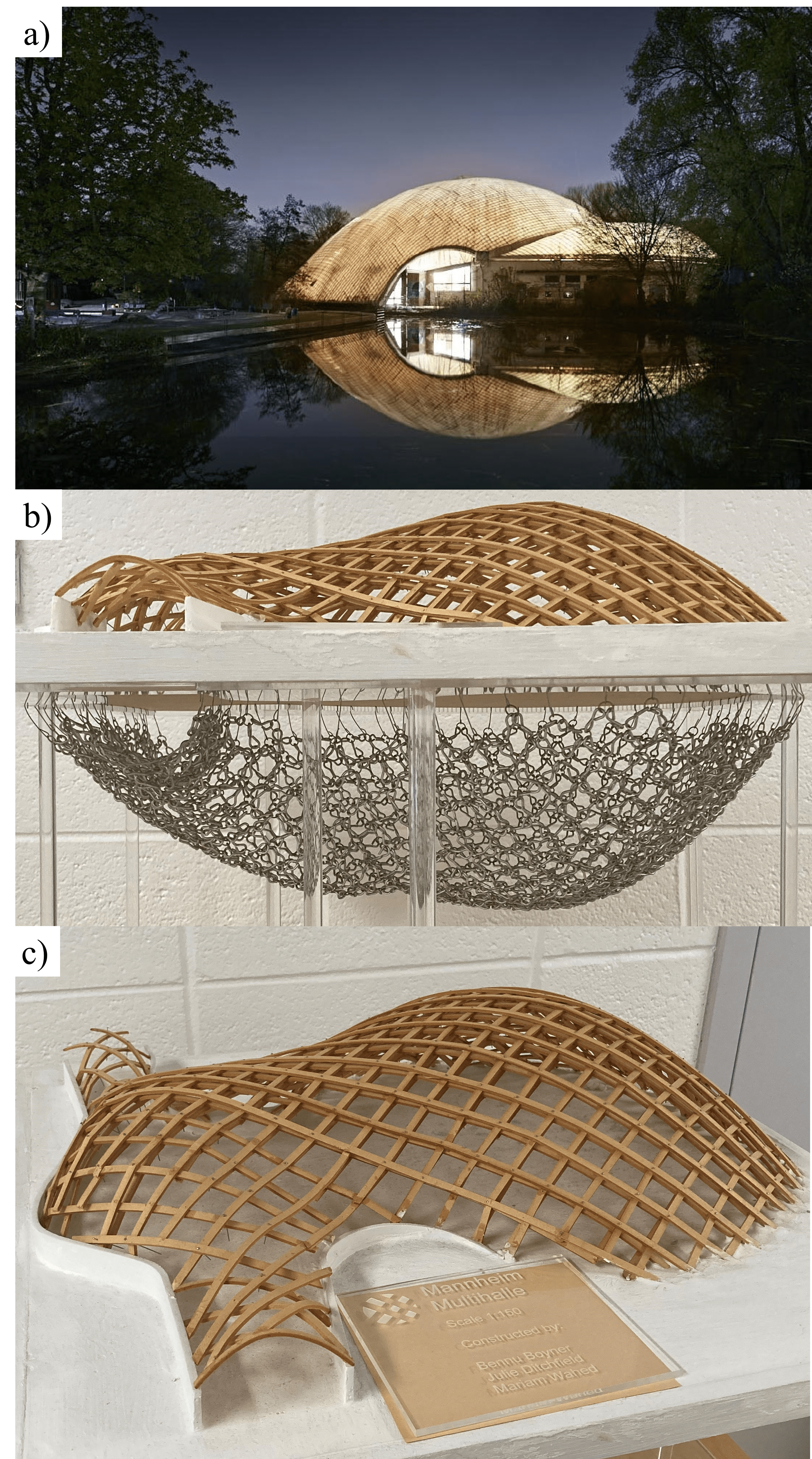}
    \caption{Mannheim Multihalle: a) Photo by Daniel Lukac; b) Model at Princeton University: the hanging model experiment; c) Model at Princeton University: the form-found gridshell}
    \label{fig:mann}
\end{figure}

Mannheim Multihalle (Figure \ref{fig:mann}) is a famous gridshell structure designed by Frei Otto via hanging model experiments.
It originally served as a temporary structure for a horticultural exhibition in Mannheim, Germany. Figure \ref{fig:mann}b-\ref{fig:mann}c shows a 1:150 model created by a group of undergraduate students at Princeton University, illustrating how this gridshell structure is realized by hanging model experiments. Inspired by the fascinating Mannheim Multihalle, here we apply the proposed framework to the shape optimization of an exhibition hall that is similar to Mannheim Multihalle.\par
\begin{figure}[h]
     \centering
     \includegraphics[width=0.5\textwidth]{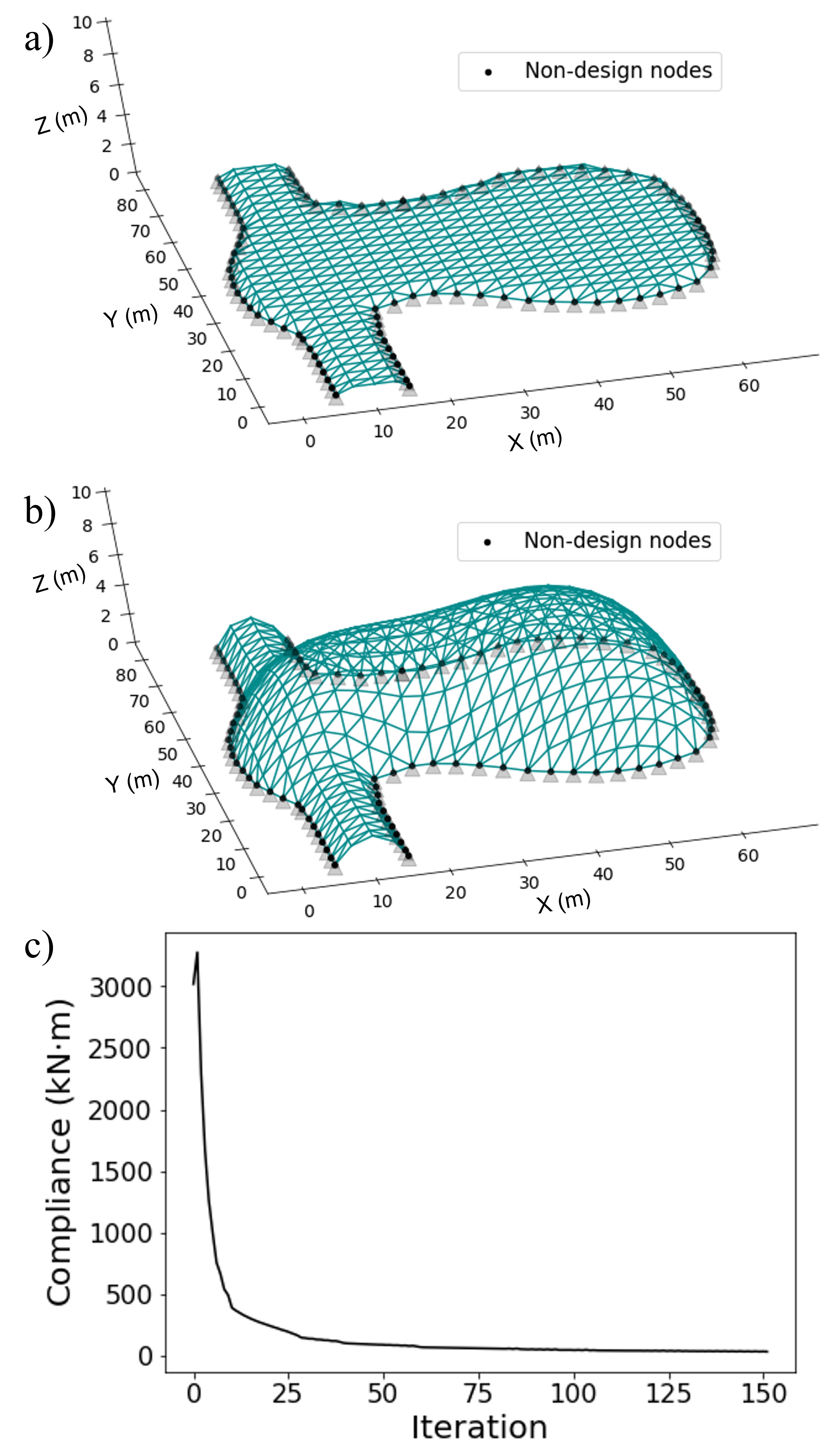}
    \caption{Mannheim Multihalle: a) initial structure of the exhibition hall, the Z-
coordinate of design nodes is 0.5m; b) optimized structure; c) Model at Princeton University: the form-found gridshell}
    \label{fig:mann_opt}
\end{figure}
The initial structure in optimization is shown in Figure \ref{fig:mann_opt}a and the plan view is similar to that of Mannheim Multihalle. The exhibition hall has a total length of 80 m and a total width of 60m. There are two entrances for the exhibition hall, located at Y=0m and Y=80m. After entering the exhibition hall, there is an exhibition space of around 2400 m$^2$. The initial structure is discretized into triangular meshes and has 440 structural nodes, which is further transformed to 1215 structural members. The loads applied to the exhibition hall is downward (-Z) nodal load at each node with a magnitude of 100 kN. The nodes on the edges of the structure are non-design nodes and they are pinned. The other 348 nodes are the design nodes with an initial height of 0.5m. In the optimization process, the design variables are thus the Z-coordinates of these design nodes. Gradient descent is selected as the optimizer for this problem. Figure \ref{fig:mann_opt}c illustrates the history of the compliance during optimization. The initial structure has a total strain energy of 3018.8 kN$\cdot$m and after 150 iterations, the total compliance is optimized to 30.3 kN$\cdot$m which is about 1\% as much as the initial compliance. The resulting geometry is shown Figure \ref{fig:mann_opt}b: between the entrances and the exhibition space, a barrel-arch shaped rooftop is found; on the top of the exhibition area, a dome-like geometry is found. The resulting geometry has a natural geometry that resembles the one shown in Figure \ref{fig:mann}c.
\subsection{Shape optimization based on B\'ezier Surface}
In the previous examples, the design variables $\bm{x}$ are the nodal Z coordinates of all the design nodes, which is the non-parametric approach to describe the structural geometry. Sometimes it may cause some issues: i) there will be too many design variables if there is an extremely fine mesh to the structure; and ii) the optimized structure may have a jagged surface, which is not ideal in architectural design for aesthetic reasons. In the examples of this subsection, we parameterize the geometry using B\'ezier Surface to reduce the number of design variables and ensure the smoothness of the structure. For a detailed introduction of B\'ezier Surface, please refer to the work by \cite{farin2014curves}. Here we only introduce the basis of B\'ezier Surface. A ($n,m$) degree B\'ezier Surface has a set of ($n+1,m+1$) control points $\bm{p}_{i,j}$ that controls the geometry where $i\in\{0,...,n\}$ and $j\in\{0,...,m\}$. The global coordinate of any point $\bm{P}$ is then a function of a set of parametric coordinate $(u,v)$:
\begin{equation}
    \bm{P}(u,v) = \Sigma_{i=0}^n\Sigma_{j=0}^{m}B_i^n(u)B_j^m(v)\bm{p}_{i,j}
\end{equation}
where $B_i^n(u)$ is a basis Bernstein polynomial:
\begin{equation}
    B_i^n(u)=\frac{n!}{i!(n-u)!}u^i(1-u)^{n-i}
\end{equation}
\begin{figure}[h]
     \centering
     \includegraphics[width=0.5\textwidth]{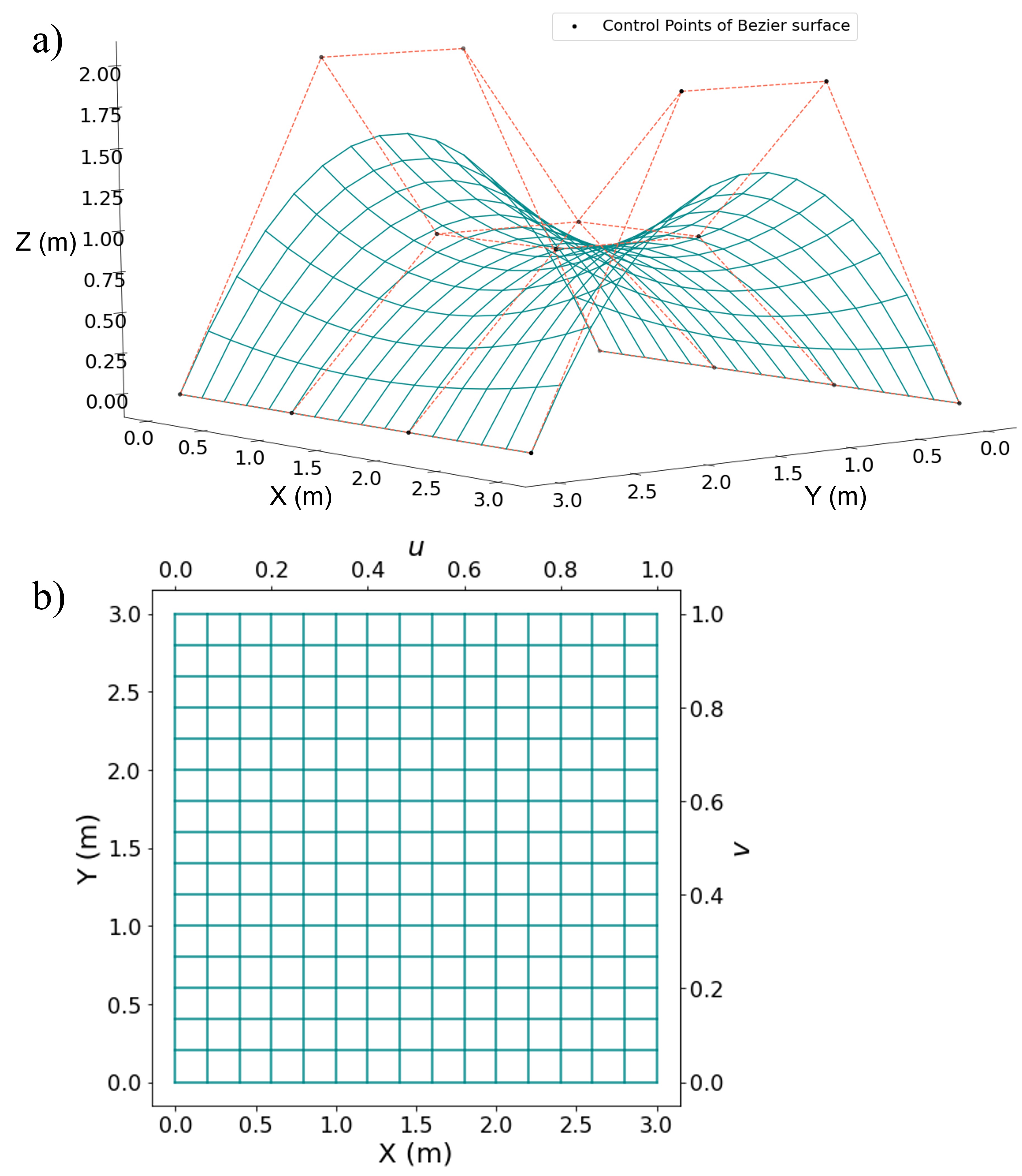}
    \caption{Geometry parameterized by B\'ezier Surface: a) A B\'ezier Surface and control points; b) Projected mesh on the X-Y plane with global coordinates (X,Y) and parametric coordinates $(u,v)$}
    \label{fig:bezier}
\end{figure}
In structural design, the parametric coordinate $(u,v)$ is determined by the mesh of the structure. An example of the geometry parameterized by B\'ezier Surface can be seen in Figure \ref{fig:bezier}. In the following examples, the design variable is the Z-coordinate $p_{z}$ of the control point $\bm{p}$. The derivatives of the objective function $C$ or of the stiffness matrix $\bm{K}$ with respect to the design variable can be easily obtained by applying the chain rule together with AD. For instance, $\frac{\partial \bm{K}}{\partial p_z} = \frac{\partial \bm{K}}{\partial \bm{P}} \frac{\partial \bm{P}}{\partial p_z}$, where the first term can be obtained from JAX-SSO and the second term can be obtained using AD. 
\subsubsection{Four-point supported free-form gridshell}
\begin{figure}[h]
     \centering
     \includegraphics[width=0.5\textwidth]{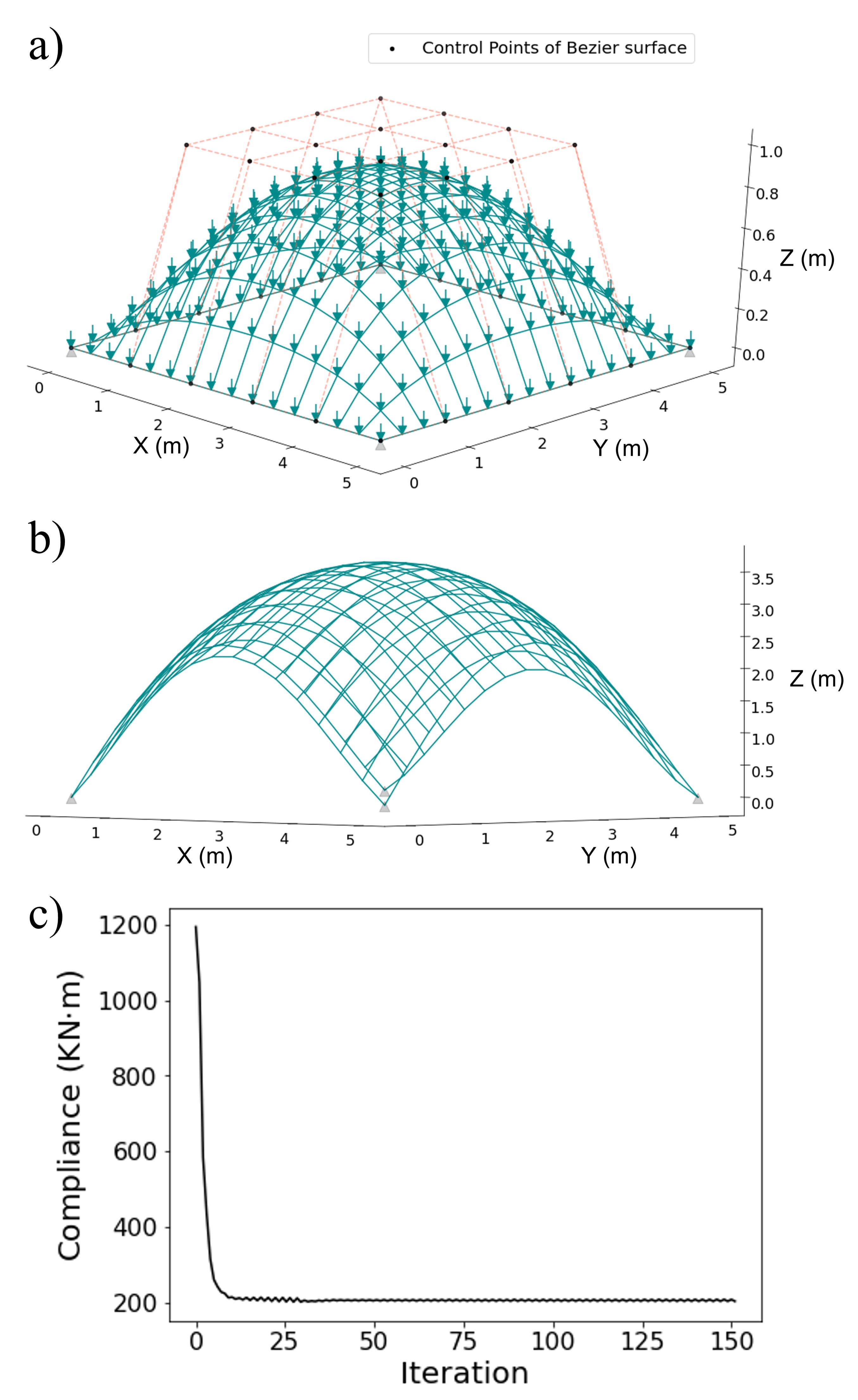}
    \caption{The optimization of a four-point supported gridshell: a) Initial structure; b) Optimized structure with gradient descent; c) Optimization history of the compliance}
    \label{fig:fourpt}
\end{figure}

The framework is applied to form-find a gridshell that has four corner-point supports (Figure \ref{fig:fourpt}), which is a classic form finding problem.
The structure (Figure \ref{fig:fourpt}a) spans 5m along the X direction and 5m along the Y direction where only the corner-nodes are pinned. The initial geometry is a dome-like geometry, which is parameterized by 36 control points of B\'ezier Surface. The Z-coordinate of the control points that are along the edges is 0m and otherwise the control points have a height of 1m. The initial geometry is ideal when all four edges are supported but intuitively it is not efficient when only the corner-nodes are constrained. The control points at the corners remain unchanged while the Z-coordinates of other control points are the design variables in the optimization process. The structure is discretized by quadrilateral mesh and there are 256 nodes in the structural system. Each node is subjected to a 500KN load along -Z axis.\par
To find the optimum geometry for this four-point supported gridshell, gradient descent is selected as the optimizer with a maximum iteration number of 150. For the initial structure, the total strain energy is 1194.2 KN$\cdot$m and after 150 iterations, the compliance converges to about 210 KN$\cdot$m (Figure \ref{fig:fourpt}c). The optimized structure's total strain energy is only 17.6 \% of the initial structure. The overall geometry of the optimized structure has positive Gaussian curvature that acts like a dome, which is similar to the initial one. However, the edge of the optimized structure changes from a straight line to an arch, which is an ideal load path to transport the nodal loads to the support by membrane forces. In addition, the resulting shape corresponds to the shape from equilibrium-state based form finding and as one can see in \cite{bletzinger2005computational}, which further validates the proposed framework.
\subsubsection{Two-edge supported free-form gridshell}
\begin{figure}[h]
     \centering
     \includegraphics[width=0.5\textwidth]{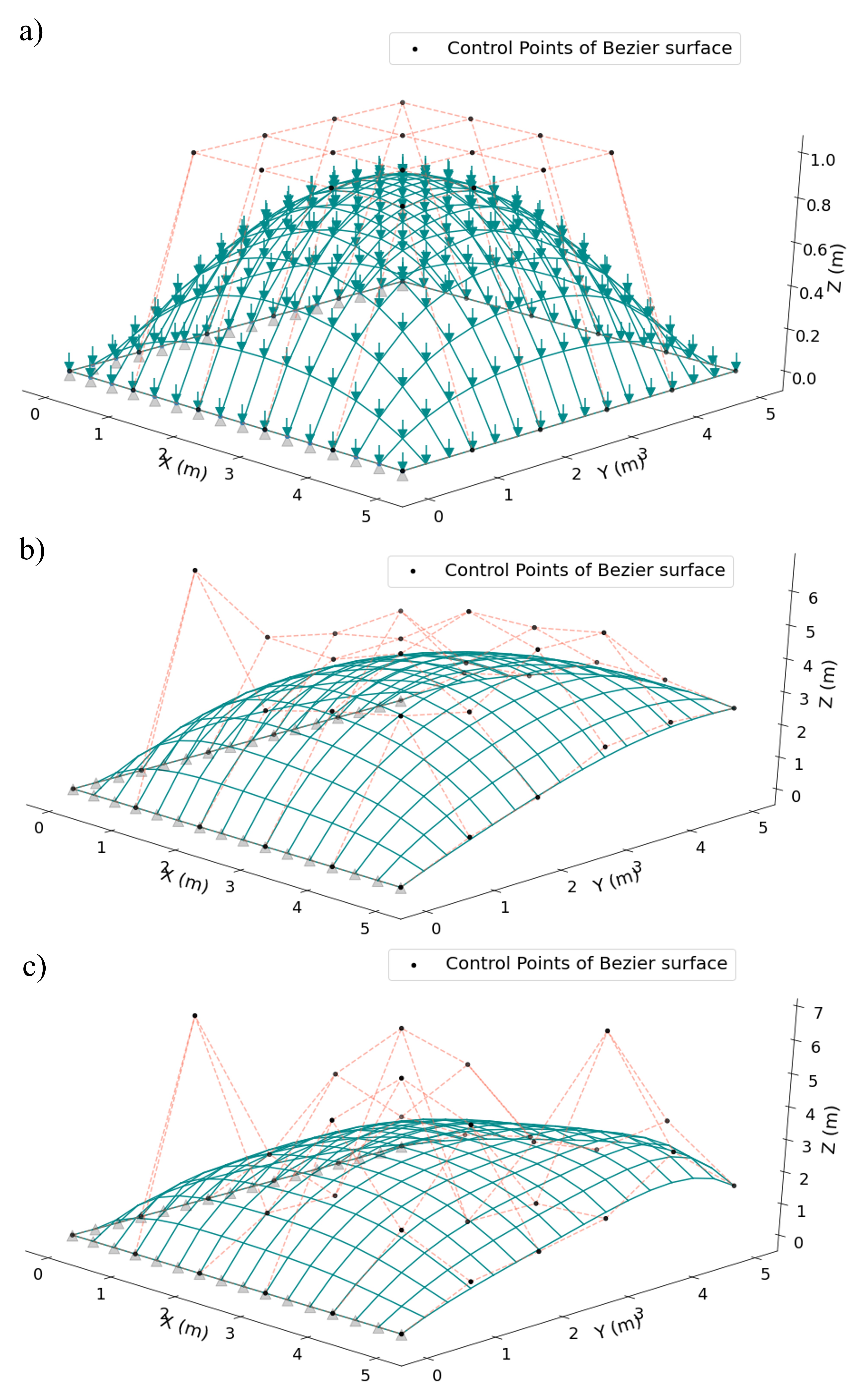}
    \caption{The optimization of a two-edge supported gridshell: a) Initial structure; b) Optimized structure without constraints, gradient descent; c) Optimized structure with constraints, SLSQP}
    \label{fig:twoedge}
\end{figure}

\begin{figure}[h]
     \centering
     \includegraphics[width=0.4\textwidth]{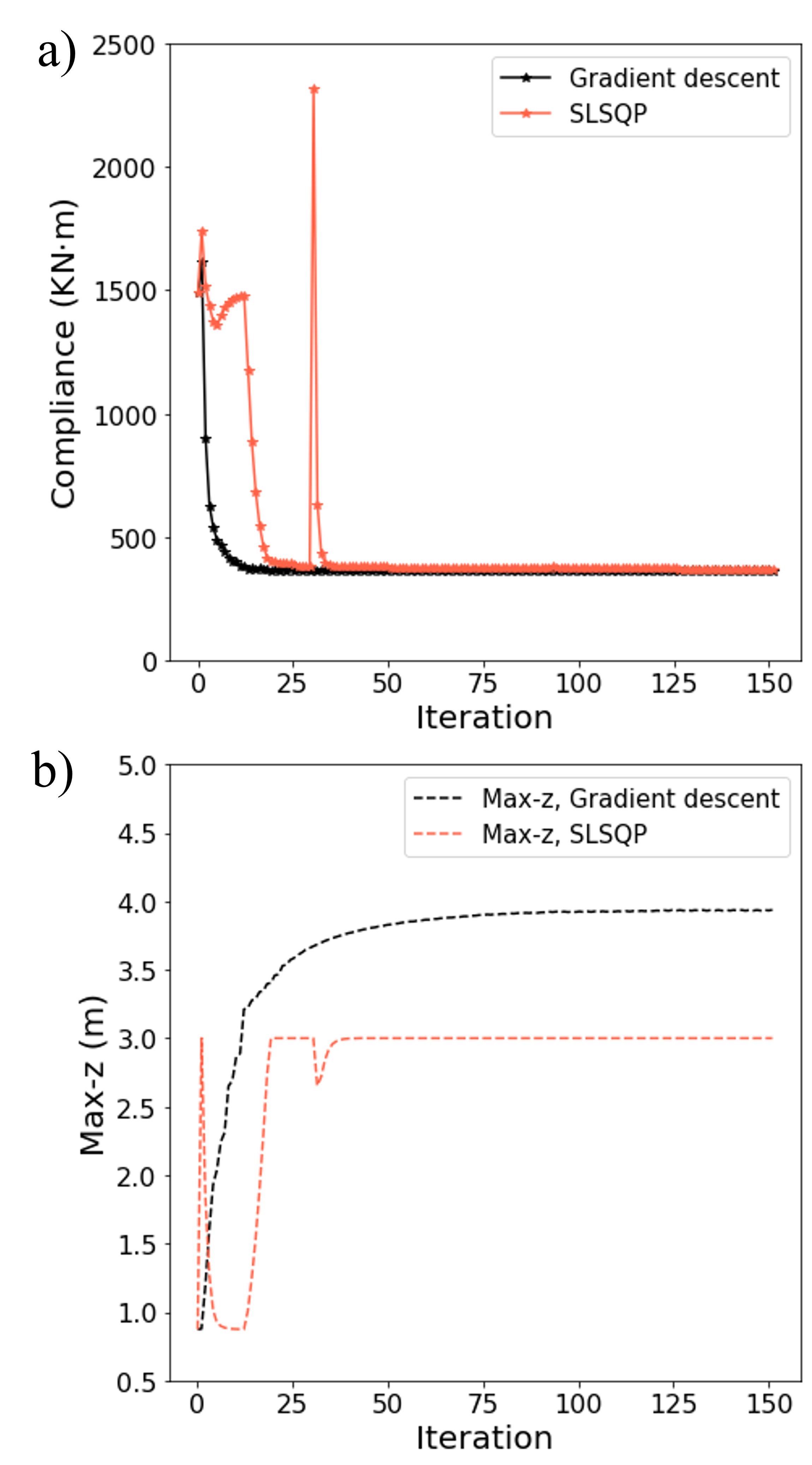}
    \caption{The optimization history of a two-edge supported gridshell: a) History of the compliance; b) History of the maximum height}
    \label{fig:twoedge_his}
\end{figure}

We then apply the proposed framework to find the optimal geometry of a gridshell that is supported along two adjacent edges (Figure \ref{fig:twoedge}).
The initial geometry, the load and the mesh are the same as the previous example. The boundary condition in this example gives a design problem of finding an optimized geometry for a canopy. Gradient descent is implemented as the optimizer and the maximum iteration number is 150. With the canopy-like boundary condition, the dome-like geometry has a total strain energy of 1487KN$\cdot$m. After 150 iterations, gradient descent reduces the total strain energy by 76\% to to 357KN$\cdot$m (Figure \ref{fig:twoedge_his}a). The optimized structure is a naturally-formed canopy (Figure \ref{fig:twoedge}b) with positive Gaussian curvature.\par
The optimized result from gradient descent is promising in terms of making efficient use of the material by minimizing the compliance. However, in structural design, constraints usually need to be applied to the shape optimization process, such as the constraint on the height of the structure. Without any constraint, the optimized structure has a maximum height of 3.933m. We now limit the maximum height to 3m: $0\text{m}\leq \text{z}\leq3\text{m}$. Mathematically, a constrained nonlinear programming problem needs to be solved. The SLSQP algorithm is selected \citep{kraft1988software} herein to find the optimal geometry when height constraints are present. Figure \ref{fig:twoedge_his}a and \ref{fig:twoedge_his}b illustrate the optimization history of the compliance and the maximum structural height during the optimization process, respectively. From SLSQP, the compliance is reduced from 1487 KN$\cdot$m to 368.8 KN$\cdot$m after 150 iterations, which is similar to the final compliance without the height constraint. However, when it comes to the maximum structural height, SLSQP successfully limits the maximum height less than 3m during the optimization while in the unconstrained optimization the height increases gradually until it finds the optimal geometry. In terms of the optimized geometries, both methods return a canopy-like geometry with a positive Gaussian curvature, which transfers the external load to the edge-supports by axial compression. In the unconstrained case (Figure \ref{fig:twoedge}b), the control points are more coherent as the height difference between adjacent control points is not significant. In contrast, to limit the maximum structural height to 3m, the height difference between adjacent control points is significant (Figure \ref{fig:twoedge}c). As can be seen in Figure \ref{fig:twoedge}c, the height of some control points is lower than others as if they are ``pulling" the structure down.
\subsubsection{Free-form canopy inspired by Carioca Wave}
\begin{figure}[h]
     \centering
     \includegraphics[width=0.5\textwidth]{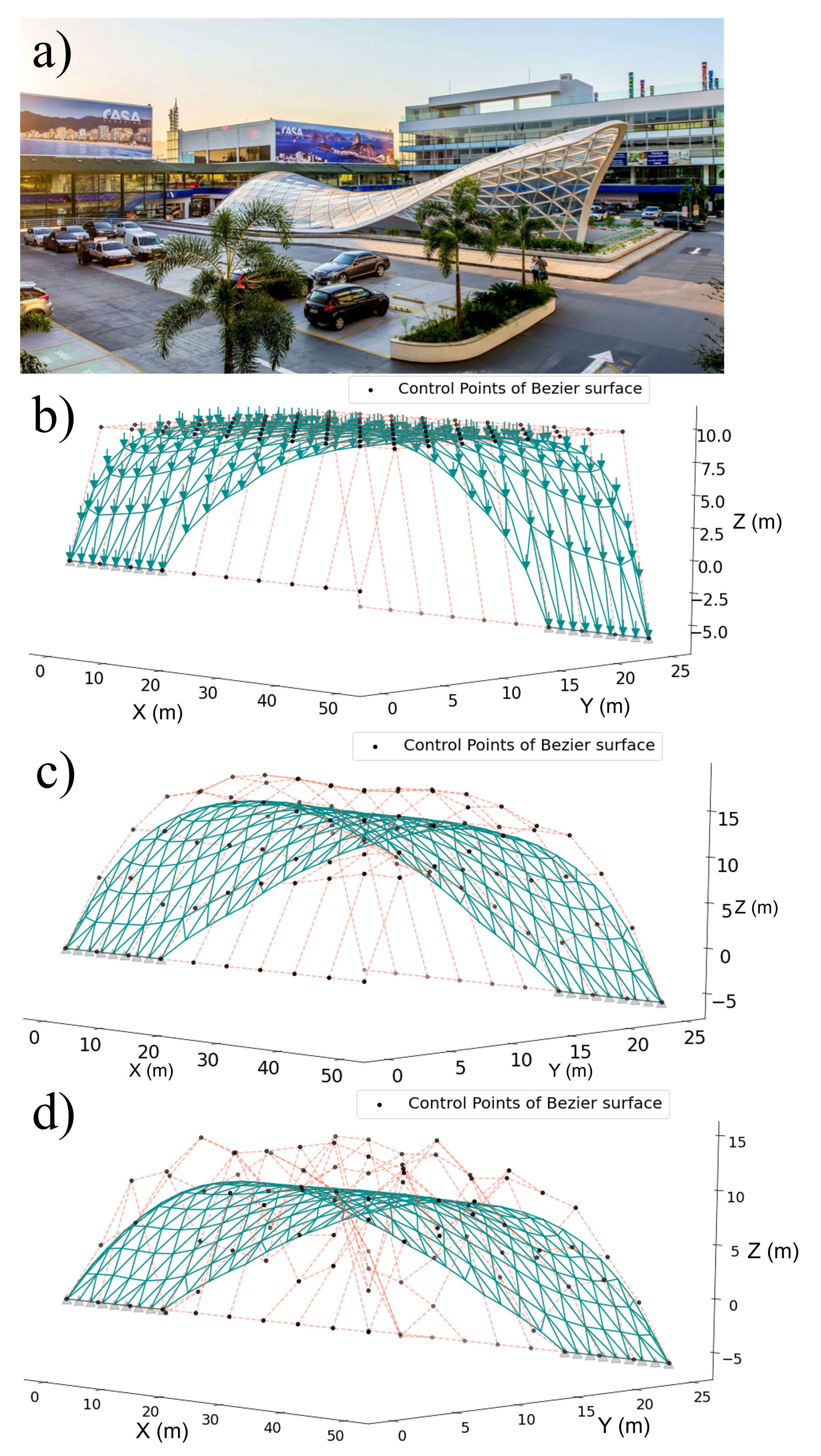}
        \caption{Form finding of a free-form canopy inspired by Carioca Wave: a) ``Carioca Wave" \textcopyright \text{ }se-austria; b) Initial strucutre; c) Optimized structure without constraints, gradient descent; d) Optimized structure with constraints, SLSQP}
    \label{fig:carioca}
\end{figure}

``Carioca Wave" \citep{helbig2014carioca} is a free-form shell canopy located in Rio de Janerio (Figure \ref{fig:carioca}a), which covers an atrium of a shopping center with its elegant geometry.
In this example, the proposed framework is applied to the shape optimization of a free-form canopy and the problem is modified from the the design of ``Carioca Wave". The initial structure is shown in Figure \ref{fig:carioca}b, where the total length of the structure is 50m and the total width is 25m. About 1/3 of the edge along the X axis is supported with pin supports. When Y=0, the edge is supported at Z=0 while the other edge is supported at Z=-6m. The initial geometry is parameterized by a B\'ezier Surface. There is a grid of 10$\times$10 control points and the height of control points are set to 10m if they are not along the supported edges. The load applied is 100kN downward (-Z) nodal load at each structural node. Two optimization problems are considered: i) unconstrained shape optimization solved by gradient descent; ii) constrained shape optimization solved by SLSQP where the maximum height and minimum height are constrained: -6m$\leq$ Z $\leq 10$m. In the optimization process, the design variables are the the Z-coordinates of the control points that are not along the supported edge.\par 
\begin{figure}[h]
     \centering
     \includegraphics[width=0.4\textwidth]{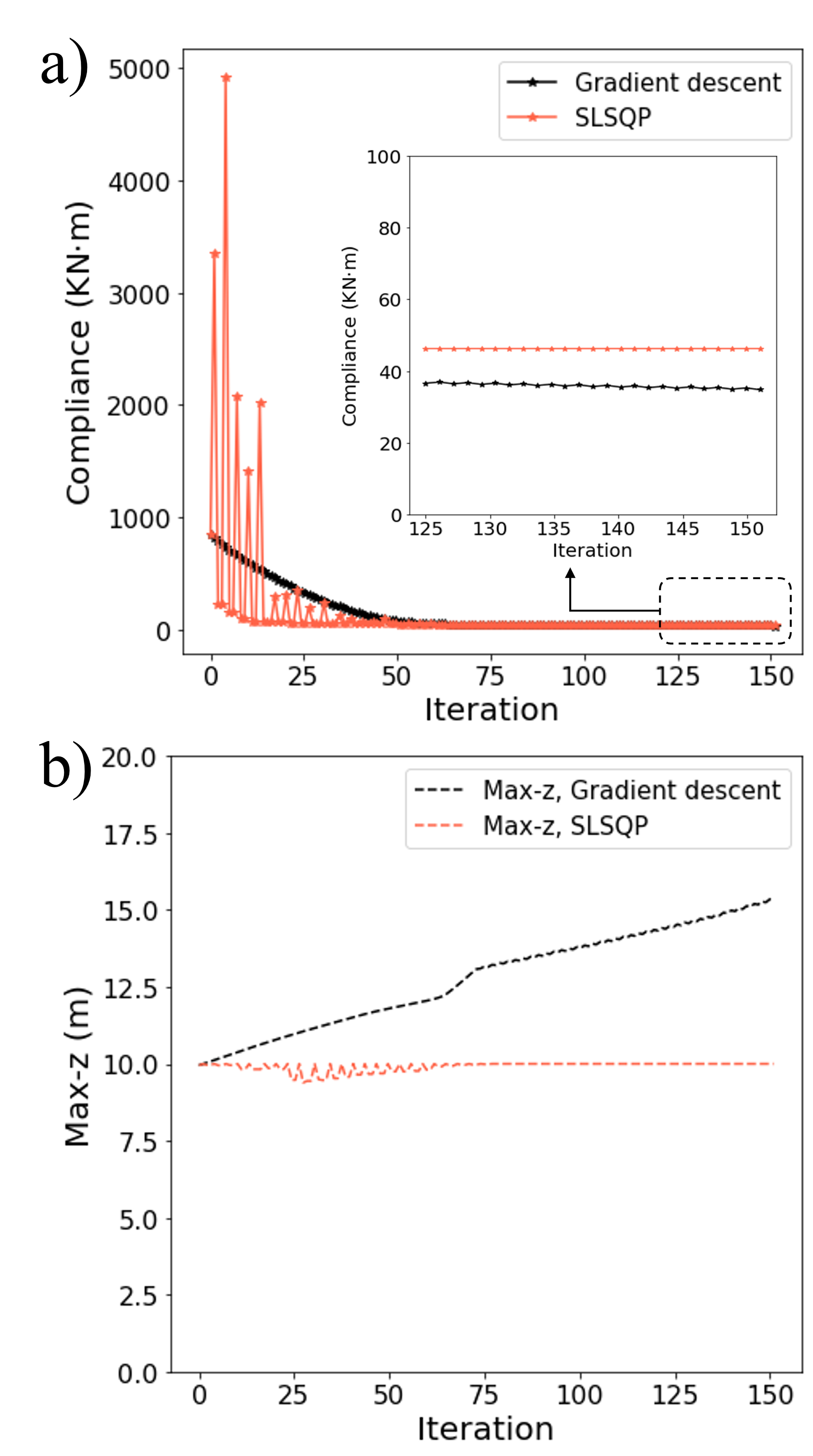}
        \caption{The optimization history of a free-form canopy inspired by Carioca Wave: a) History of the compliance; b) History of the maximum height}
    \label{fig:carioca_his}
\end{figure}
\begin{figure}[h]
\centering
\includegraphics[width=1.1\linewidth]{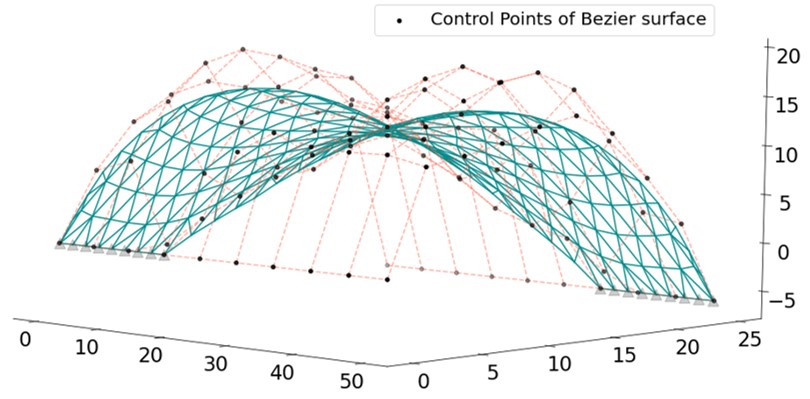}
\caption{Optimized structure with reduced bending stiffness}
\label{fig:carioca_gd_small}
\end{figure}
The optimized shape using gradient descent can be seen in Figure \ref{fig:carioca}c and the optimized geometry using SLSQP is illustrated in Figure \ref{fig:carioca}d.
A history of the compliance and the maximum structural height can be seen in Figure \ref{fig:carioca_his}a and Figure \ref{fig:carioca_his}b, respectively. Different from the initial structure that all the design control points have a height of 10m, the optimized geometry in both constrained and unconstrained problems forms a barrel arch shape that carries the force to the supports by axial force. The initial structure has a compliance of 842 kN$\cdot$m and after 150 iterations, the final structure obtained from gradient descent has a compliance of 34.8 kN$\cdot$m, decreasing by 96\%. As can be seen in Figure \ref{fig:carioca_his}b, the maximum height gradually increases to 15.3m in the unconstrained problem while in the constrained problem, the maximum height is successfully limited to less than 10m by SLSQP. In terms of the compliance, the optimized compliance by SLSQP is 46.5 kN$\cdot$m, which is 33\% higher than the compliance in the unconstrained problem. The difference in the compliance shows a trade off between the objective and the constraints. \par
Compared to the Carioca Wave, the form-found geometries in the above examples are not doubly curved. One of the reasons is due to the properties of the structural elements used, especially the bending stiffness. To investigate how the bending stiffness of structural elements influences the optimized geometry, an additional unconstrained optimization problem is conducted in which the moment of inertia $I_y$ and $I_z$ are reduced to 10\% of the original values shown in Table \ref{table:members}. Figure \ref{fig:carioca_gd_small} shows the optimized structure after 150 iteration by gradient descent. With the decrease of the bending stiffness, the resulting structure is doubly curved and has a negative Gaussian curvature, similar to the Carioca Wave. The final geometry shows that by having smaller bending stiffness in structural members, the load-resisting mechanism is transformed to a combination of compression and tension, manifested by the negative Gaussian curvature. 
\section{Conclusions}\label{sec:con}
To facilitate efficient and rigorous gradient-based structural shape optimization for structural design, a framework based on automatic differentiation (AD), accelerated linear algebra (XLA) and the adjoint method is proposed herein. To the best of the authors' knowledge, this is the first implementation of the high-performance-computing (HPC) library JAX for FEA-based structural shape optimization. The features of this framework are highlighted as follows.
\begin{itemize}
    \item JAX-enabled AD and the adjoint method for gradient calculation: in gradient-based structural shape optimization, the gradient calculation is of great significance and it is often the bottleneck. Traditionally, manual derivation or numerical differentiation is used. The former is time-consuming and error-prone while the latter is also time-consuming and suffers from truncation and round-off errors. The implementation of AD makes this hard yet important step efficient and accurate. Moreover, the employment of the adjoint method further reduces the complexity of the gradient calculation of the objective function.
    \item JAX-enabled XLA and vectorization for fast gradient calculation: in the JAX-SSO Python package that we developed for this framework, the XLA feature in JAX boosts the gradient evaluation. The derivative evaluation is conducted simultaneously for each structural element by vectorization, which is further accelerated by just-in-time compilation through XLA. This also enables GPU-acceleration for even faster calculation. 
    \item Coherence between structural analysis and optimization: Different from equilibrium-state based form finding methods in which the structural members used are different from the members used in the actual design and construction process, the optimization process in our framework is informed by the properties of the actual structural elements. In this framework, the FEA is conducted alongside with the optimization process and the structural response is obtained once the optimization is done. By applying this framework, there will be less design iterations as the optimization process is coherent with the analysis using FEM, which is time-saving for the designers.
    \item Broader fields for application: while equilibrium-state based form finding is mainly for the design of lightweight structures with specific boundary conditions, this framework can be applied to more structural systems. For example, by using the actual structural elements that have bending stiffness, form-finding for cantilever-like structures will be possible, which is hard to realize by equilibrium-state based form finding methods \citep{ding2017new}.
\end{itemize} 
The proposed framework is applied to several form-finding test cases, including  arches and free-form shells. The optimization results verify the validity of the proposed framework and the JAX-SSO tool. Two geometric description approaches are used. The non-parametric approach is used to form-find the arches and the exhibition hall inspired by Mannheim Multihalle. The parametric approach is used to form-find several free-form gridshells with different boundary conditions and constraints are applied to some cases. The advantage of the parametric geometric description is that it ensures a smooth geometry while reducing the number of design variables. The advanatge of the non-parametric approach is that it will find a solution closer to the optimum when compared to the parametric approach. In terms of the optimizers, gradient descent works well with unconstrained problems and SLSQP works well for the constrained problems in this work. However, it has to be pointed out that both algorithms may stuck at local minimum so future work should be conducted to see how the initial geometry influences the optimized geometry. In addition, the properties of the structural elements will influence the optimized geometry, which shows the necessity of implementing the shape optimization approach that is informed by the properties of structural elements. \par 
While this paper illustrates the validity and efficiency of the proposed framework for structural shape optimization, future research and development still need to be conducted to further facilitate the employment of structural shape optimization based on optimization theory and FEA:
\begin{itemize}
    \item Development of the JAX-SSO package: the JAX-SSO package needs to be extended to support more structural elements, such as shell elements. In addition, while calculating $\frac{\partial \bm{K}}{\partial \bm{x}}$ is fast in JAX-SSO, the time of the optimization now mainly depends on the time of the FE analysis. The JAX-SSO package is expected to be extended into a differentiable FEA solver boosted by XLA.
    \item Comparison between different optimizers: while both gradient descent and SLSQP work well in this work, a systematic comparison between commonly used optimization algorithms should be conducted to inform the designers which algorithms should be selected for their design problems.
    \item Comparison between optimization-theory-and-FEA based shape optimization and equilibrium-state based form-finding: very limited work offers insights on the comparison between these two approaches. As such, future work is needed to systematically evaluate their capability and weakness. Aspects such as the time needed for optimization, the optimization quality, and the range of applicable design problems should be investigated. In this way the designers are then able to choose a more suitable tool once they identify the design problems.
    \item A combination of equilibrium-state based form finding and optimization-theory-and-FEA based shape optimization: while the work by \cite{bletzinger2005computational} has shown the potentials of a merged method consisting of equilibrium-state based form finding methods and optimization-theory-and-FEA based shape optimization, more work should be conducted to evaluate the efficiency and capability of this merged approach.
    \item AD for structural shape optimization with nonlinearity: in the proposed framework, linear FEA is conducted. However, when large deformation is present and when material is damaged, the geometrical linearity and the material linearity do not hold anymore. Future work should be conducted to tackle structural shape optimization with nonlinear effects using AD. 
\end{itemize}
Nevertheless, this paper shows the potentials of implementing AD, XLA and the adjoint method for efficient shape optimization. In addition, with the assistance of HPC, the shape optimization process can be further boosted, enabling designers to deal with design problems with high complexity.
\backmatter

\bmhead{Acknowledgments}
The author would like to thank his advisor Professor Maria Garlock and cohorts at Princeton University for the insightful discussions.

\section*{Declarations}
\subsection*{Declaration of Competing Interest}
The authors declare that they have no known competing financial interests or personal relationships that could have appeared to influence the work reported in this paper.
\subsection*{Code availability}
The code for this paper is available online: \href{https://github.com/GaoyuanWu/JaxSSO}{https://github.com/GaoyuanWu/JaxSSO}
\subsection*{Authors' contributions}
\textbf{Gaoyuan Wu}: Conceptualization, Methodology, Software, Validation, Formal analysis, Writing – original draft, Writing – review \& editing.

\begin{appendices}

\section{Code example: derivative calculation using JAX for customized Python objects}\label{app:app_1}
The code snippet (Code \ref{code:toy_jax}) illustrates how the JAX methods work by presenting a simple problem: calculating how the length of a 2D line changes with the nodal coordinates, i.e., the gradient of the length with respect to the nodal coordinates of the line. Firstly, a new class that represents a line in 2D world is defined and its attributes are the nodal coordinates. The {\ttfamily line()} class is registered to JAX by {\ttfamily @register\_pytree\_node\_class}, {\ttfamily tree\_flatten} and {\ttfamily tree\_unflatten}. A function called {\ttfamily L()} is firstly defined, outputting the distance between two 2D points. Note that {\ttfamily jax.numpy} is used in the function {\ttfamily L()} for calculating distance instead of {\ttfamily numpy}, which is of great significance because we need to trace the operations so that the derivatives can be obtained through automatic differentiation.  An external function is then defined for {\ttfamily line()} objects to calculate the derivatives of its length with respect to nodal coordinates. 100 lines with random coordinates are then created. Lastly, we implement {\ttfamily jit} and {\ttfamily vmap} to compile the function for derivative calculation so that the derivative calculation is sped up.
\begin{lstlisting}[language=Python, label={code:toy_jax}, caption=An example of derivative calculation using JAX for customized Python objects]
#Import packages needed
import numpy as np
import jax.numpy as jnp
from jax import jit, vmap, jacfwd
from jax.tree_util import register_pytree_node_class


#Register line() class as JAX's Pytrees
@register_pytree_node_class #Decorator
class line():
  '''
  A class for 2D line
  '''
  def __init__(self,x1,x2,y1,y2):
    self.x1 = x1 # x of the first node
    self.x2 = x2 # x of the second node
    self.y1 = x1 # y of the first node
    self.y2 = x2 # z of the second node

  #Tell JAX how to 
  #flatten the attributes of line() object
  def tree_flatten(self):
    children = (self.x1,self.x2,self.y1,self.y2)
    aux_data = None
    return (children, aux_data)

  #Tell JAX how to 
  #unflatten the attributes of line() object
  @classmethod
  def tree_unflatten(cls, aux_data, children):
    return cls(*children)

    
#The distance between two points in 2D
def L(x1,x2,y1,y2):
  return jnp.sqrt(((x1-x2)**2+(y1-y2)**2)) 

#Function for line() objects:
#calculates the sensitivity of the length
#w.r.t. nodal coordinates
def line_sens(a_line):
  x1 = a_line.x1
  x2 = a_line.x2
  y1 = a_line.y1
  y2 = a_line.y2
  
  #jacfwd for derivative calculation
  return jacfwd(L,argnums=(0,1,2,3))(x1,x2,y1,y2)

#Create 100 lines
coords_random = np.random.randn(100,4) #random lines
x1 = coords_random[:,0]
y1 = coords_random[:,1]
x2 = coords_random[:,2]
y2 = coords_random[:,3]
lines = line(x1,y1,x2,y1) #100 lines

#Compile 'line_sens' function
#using vmap and jit
#for fast implementation
compiled_line_sens = vmap(jit(line_sens))

#Calculate the sensitivity
sens = compiled_line_sens(lines)

\end{lstlisting}

\end{appendices}


\bibliography{sn-bibliography}


\end{document}